\documentclass[aps,prd,10pt,amsmath,amssymb,twocolumn,nofootinbib,groupedaddress,superscriptaddress,floatfix]{revtex4-2}

\usepackage{graphicx}
\usepackage{dcolumn}
\usepackage{bm}
\usepackage{microtype}
\usepackage{xcolor}
\definecolor{navyblue}{rgb}{0.0, 0.0, 0.5}
\usepackage{hyperref}
\hypersetup{
    colorlinks=true,
    linktoc=all,
    allcolors=navyblue
}
\usepackage[capitalize]{cleveref}
\crefname{section}{Sec.}{Secs.}
\Crefname{section}{Section}{Sections}
\crefname{appendix}{App.}{Apps.}
\Crefname{appendix}{Appendix}{Appendices}
\crefname{figure}{Fig.}{Figs.}
\Crefname{figure}{Figure}{Figures}
\creflabelformat{equation}{#2#1#3}

\usepackage{color,environ}
\usepackage{fontawesome5}
\definecolor{orcidlogocol}{rgb}{0.65, 0.807, 0.223}
\newcommand{\orcid}[1]{$\,$\href{https://orcid.org/#1}{\textcolor{orcidlogocol}{\footnotesize\faOrcid}}}

\graphicspath{{./}{./}}

\newcommand*\gen[1]{\langle #1 \rangle}

\newcommand*\te[1]{\text{#1}}

\newcommand*\p[1]{\left(#1\right)}
\newcommand*\ps[1]{\left[#1\right]}

\newcommand*\f[2]{\frac{#1}{#2}}

\newcommand{\ths}{\theta_S}
\newcommand{\thl}{\theta_L}

\newcommand{\phs}{\phi_S}
\newcommand{\phl}{\phi_L}

\newcommand{\calF}{\mathcal{F}}
\newcommand{\gagg}{g_{a \gamma \gamma}}

\newcommand{\eV}{\; \mathrm{eV}}
\newcommand{\Mpl}{M_\mathrm{pl}}

\newcommand{\overbar}[1]{\mkern 1.25mu\overline{\mkern-1.25mu#1\mkern-1.25mu}\mkern 1.25mu}
\newcommand{\ud}{\mathrm{d}}
\newcommand{\dd}[2]{\frac{\ud {#1} }{\ud {#2}}}
\newcommand{\tnl}{t_{\mathrm{nl}}}
\newcommand{\rhotot}{\bar{\rho}_{\mathrm{tot}}}

\NewEnviron{eq}{%
\begin{align}\begin{split}
  \BODY
\end{split}\end{align}
}

\usepackage[normalem]{ulem}
\usepackage{mathtools}  

\makeatletter
\newcommand{\amarki}{\faCoffee}
\newcommand{\amarkii}{\faRebel}
\newcommand{\amarkiii}{\faPaperPlane[regular]}
\newcommand{\amarkiv}{\faBeer}
\def\@fnsymbol#1{{\ifcase#1\or \amarki\or \amarkii\or \amarkiii\or \amarkiv \else\@ctrerr\fi}}
\makeatother

\begin{document}
\raggedbottom


\title{Nonperturbative structure in coupled axion sectors \\ and implications for direct detection}

\author{David Cyncynates\orcid{0000-0002-2660-8407}}
\email{davidcyn@uw.edu}
\affiliation{Department of Physics, University of Washington, Seattle, Washington 98195, USA}
\affiliation{Stanford Institute for Theoretical Physics, Stanford University, Stanford, California 94305, USA}

\author{Olivier Simon\orcid{0000-0003-2718-2927}}
\email{osimon@stanford.edu}
\affiliation{Stanford Institute for Theoretical Physics, Stanford University, Stanford, California 94305, USA}

\author{Jedidiah O. Thompson\orcid{0000-0002-7342-0554}}
\email{jedidiah@stanford.edu}
\affiliation{Stanford Institute for Theoretical Physics, Stanford University, Stanford, California 94305, USA}

\author{Zachary J. Weiner\orcid{0000-0003-1755-2277}}
\email{zweiner@uw.edu}
\affiliation{Department of Physics, University of Washington, Seattle, Washington 98195, USA}

\date{\today}

\begin{abstract}
Pairs of misalignment-produced axions with nearby masses can experience a nonlinear resonance that leads to enhanced direct and astrophysical signatures of axion dark matter.
In much of the relevant parameter space, self-interactions cause axion fluctuations to become nonperturbative and to collapse in the early Universe.
We investigate the observational consequences of such nonperturbative structure in this ``friendly axion'' scenario with $3+1$ dimensional simulations.
Critically, in a substantial fraction of parameter space we find that nonlinear dynamics work to equilibrate the abundance of the two axions,
making it easier than previously expected to experimentally confirm the existence of a resonant pair.
We also compute the gravitational wave emission from friendly axion dark matter; while the
resulting stochastic background is likely undetectable for axion masses above $10^{-22} \,
\text{eV}$, the polarization of the cosmic microwave background does constrain possible hyperlight,
friendly subcomponents.
Finally, we demonstrate that dense, self-interaction--bound oscillons formed during the period of strong nonlinearity are driven by the homogeneous axion background, enhancing their lifetime beyond the in-vacuum expectation.
\end{abstract}

\maketitle

\section{Introduction} \label{sec:intro}

Axions are some of the best-motivated extensions to the Standard Model (SM).
The simplest such extension, the QCD axion, was originally proposed to solve the strong $CP$
problem~\cite{Peccei:1977hh,Peccei:1977ur,Weinberg:1977ma,Wilczek:1977pj}, but it has
since been realized that axions are common in many theories beyond the SM
(BSM)~\cite{Witten:1984dg,Banks:1996ss,Svrcek:2006yi}.
One particularly important example is string theory, which generically predicts a large number of light axions coupled weakly to the SM~\cite{Arvanitaki:2009fg}.
The possibility of such a ``string axiverse'' is of particular interest because it offers a
potential low-energy window into extremely high-energy physics.

The simplest nonthermal production mechanism for a cosmological abundance of axions is the
misalignment mechanism~\cite{Preskill:1982cy,Abbott:1982af,Dine:1982ah,Turner:1983he,Marsh:2015xka}.
Any axion with a mass lighter than the Hubble scale during inflation would be seeded in an
approximately homogeneous state displaced from the vacuum.
It would then remain frozen at this ``misaligned'' field value until the expansion rate drops
below its mass, at which point it begins to coherently oscillate about the minimum of its potential.
Barring substantial sources of isocurvature, axions have large-scale density perturbations that track
the adiabatic fluctuations also seeded during inflation and are thus a viable candidate for the
observed dark matter (DM) or a subcomponent thereof.

An axion's potential is generically nonlinear, but at late times all axions with a mass $m$ much
larger than the present-day Hubble rate ($m \gg H_0$) oscillate near the bottom of their potential
and may be treated as free, massive fields.
This is not, however, a valid assumption at early times, and it has become increasingly apparent that nonlinearities in an axion's potential can have an outsized impact on many late-time observables (see, e.g., Refs.~\cite{Daido:2015bva,Kitajima:2018zco,Arvanitaki:2019rax,Co:2019jts,Cyncynates:2021xzw,Co:2020dya, Eroncel:2022vjg,Eroncel:2022efc}).
If the dark matter comprises a single axion, these early-time dynamics can strongly enhance structure on scales that enter the horizon when the Hubble rate $H$ is approximately the axion mass $m$~\cite{Arvanitaki:2019rax}.

More generally, a string axiverse may consist of many axions interacting with each other through a joint potential, and recent work has shown that when any two of these have similar masses (within a factor of roughly 2) a new type of efficient, resonant energy transfer is possible~\cite{Cyncynates:2021xzw}.
This mechanism, dubbed ``friendship'' due to the necessary mild coincidence of masses, transfers
energy from an axion with a high decay constant to one with a lower decay constant.
Since an axion's couplings to the SM are generically inversely proportional to its decay constant, the mechanism boosts the abundance of the more strongly coupled axion.
In other words, friendly axion dark matter can be significantly more visible to direct detection
experiments than would be expected for either axion individually.

In this paper, we follow up on the work of Ref.~\cite{Cyncynates:2021xzw} with a suite of $3+1$
dimensional numerical simulations, corroborating its findings and extending the results to the
strongly nonlinear regime.
As anticipated in that work, large spatial inhomogeneities significantly modify the results of a
homogeneous analysis.
Nonperturbative fluctuations collapse into dense \textit{oscillons}, nontopological field
configurations bound by self-interactions~\cite{kudryavtsev1975solitonlike,Makhankov:1978rg,Gleiser:1993pt,Kolb:1993hw,Salmi:2012ta,Amin:2011hj,Kawasaki:2019czd,Olle:2020qqy,Zhang:2020bec,Cyncynates:2021rtf}.
The oscillons quench the resonant amplification and mediate energy transfer between the friendly
pair, leading to approximate energy density equipartition over a broad range of parameters.
In contrast to expectations from a homogeneous analysis, the enhanced visibility of one axion
therefore does \textit{not} come at the expense of the other's detectability.
In sum, nonlinear dynamics make the friendly axion model both more predictive (by being less
parameter-dependent) and more identifiable (because both axions would be detectable).

This paper is divided as follows.
In \cref{sec:review} we review the friendly axion model and results within the spatially homogeneous approximation.
\Cref{sec:results} presents the extension of these results into the nonlinear regime using numerical
simulations, with a primary focus on the late-time abundance as relevant to direct detection
experiments.
We also investigate gravitational wave signatures in these scenarios, which, while not promising if
the friendly axions make up all of the dark matter, are relevant for hyperlight subcomponents.
Finally, we study a novel driving effect in which oscillons resonantly siphon energy from the
axion background, parametrically enhancing their lifetime.
We conclude in \cref{sec:discussion}, putting this work into the broader context of the
landscape of nonlinear axion models.
For completeness and ease of readability, we relegate an extended discussion of methodology and
additional results to the appendices.
\Cref{app:numerical-details} enumerates the system of evolution equations and the details of
our numerical implementation, and \cref{app:oscillons} expands upon our discussion of bound axion states.

\section{Review of friendly axions} \label{sec:review}

As a concrete and illustrative model, Ref.~\cite{Cyncynates:2021xzw} focuses on a simple two-axion
potential with two instanton contributions:\footnote{
    Throughout, we work in units where $\hbar = c = 1$.
    We also define the reduced Planck mass
    $\Mpl = 1 / \sqrt{8 \pi G} \approx 2.44 \times 10^{18} \, \mathrm{GeV}$.
}
\begin{align} \label{eq:twoAxionPotentialPhi}
    \begin{split}
    V(\phi_S,\phi_L)
    &= \Lambda_1^4 \left[ 1 - \cos \left( \frac{\phs}{f_S} + \frac{\phl}{f_L} \right) \right] \\
        &\hphantom{{}={}}
        + \Lambda_2^4 \left( 1 - \cos \frac{\phl}{f_L} \right).
    \end{split}
\end{align}
The canonically normalized axion field variables $\phi_S$ and $\phi_L$ are naturally recast as
angular variables via the definition $\ths \equiv \phs / f_S$ and $\thl \equiv \phl / f_L$.
Redefining $\Lambda_1^4 \equiv m^2 f^2$ and $\Lambda_2^4 \equiv \mu^2 m^2 \calF^2 f^2$,
the axion masses are\footnote{
    The interaction-basis axions $\phi_S$ and $\phi_L$ are not exact mass eigenstates, making this
    definition ambiguous.
    For $\calF \gg 1$, the distinction between the two bases is small, and so we often neglect the distinction in our heuristic discussions.  The effect is not, however, quantitatively
    negligible for all parts of the parameter space we consider, and it is always included in our results.
} $m_S = m$ and $m_L = \mu m_S$ and their decay constants are $f_S = f$ and $f_L = \calF f$,
respectively.
In terms of these variables, \cref{eq:twoAxionPotentialPhi} takes the form
\begin{align} \label{eq:twoAxionPotential}
    \begin{split}
    V( \thl , \ths )
    &= m^2 f^2 \Big[
        \left( 1 - \cos \left( \ths + \thl \right) \right) \\
        &\hphantom{{}={}  m^2 f^2 \Big[}
        + \mu^2 \calF^2 \left( 1 - \cos \thl \right) \Big].
    \end{split}
\end{align}
We focus on the range $\calF > 1$ where $f_S < f_L$, and we
refer to $\phs$ and $\phl$ as the ``short'' and ``long'' axion respectively in reference to the size
of their decay constants.
(The regime with $f_S > f_L$ does not exhibit nonlinear resonances.)
The short and long axions then form a ``friendly pair'' when $0.7 \lesssim \mu < 1$,
corresponding to an $\mathcal{O}(1)$ coincidence in their masses.
While \cref{eq:twoAxionPotentialPhi} might represent a subsector of a much larger axiverse, the
dynamics of the friendly pair of interest are insensitive to possible couplings to other axions
barring additional coincidences in mass.
Namely, only the relative frequency of coupled oscillators determines the efficiency of energy transfer between them, so the actual instanton scales $\Lambda_i$ and decay constants $f_i$ matter only insofar as they (together) determine the axion masses.

In the early Universe, the misalignment mechanism initializes each axion at an approximately
spatially homogeneous value away from the late-time minimum; a natural assumption, barring anthropic
and other considerations, is that $\theta_I(t_\text{initial}) = \mathcal{O}(1)$, where the capital
index $I$ runs over axion flavors.
The axions remain frozen at their misaligned values until the Hubble rate $H$ drops below their
masses.
Since the two axion masses are comparable, the long axion initially has $\mathcal{O}(\calF^2)$ times
more energy than the short axion.
In the absence of couplings between the axions, this imbalance would persist to their present-day
abundance.

\begin{figure}
    \centering
    \includegraphics[width = \columnwidth]{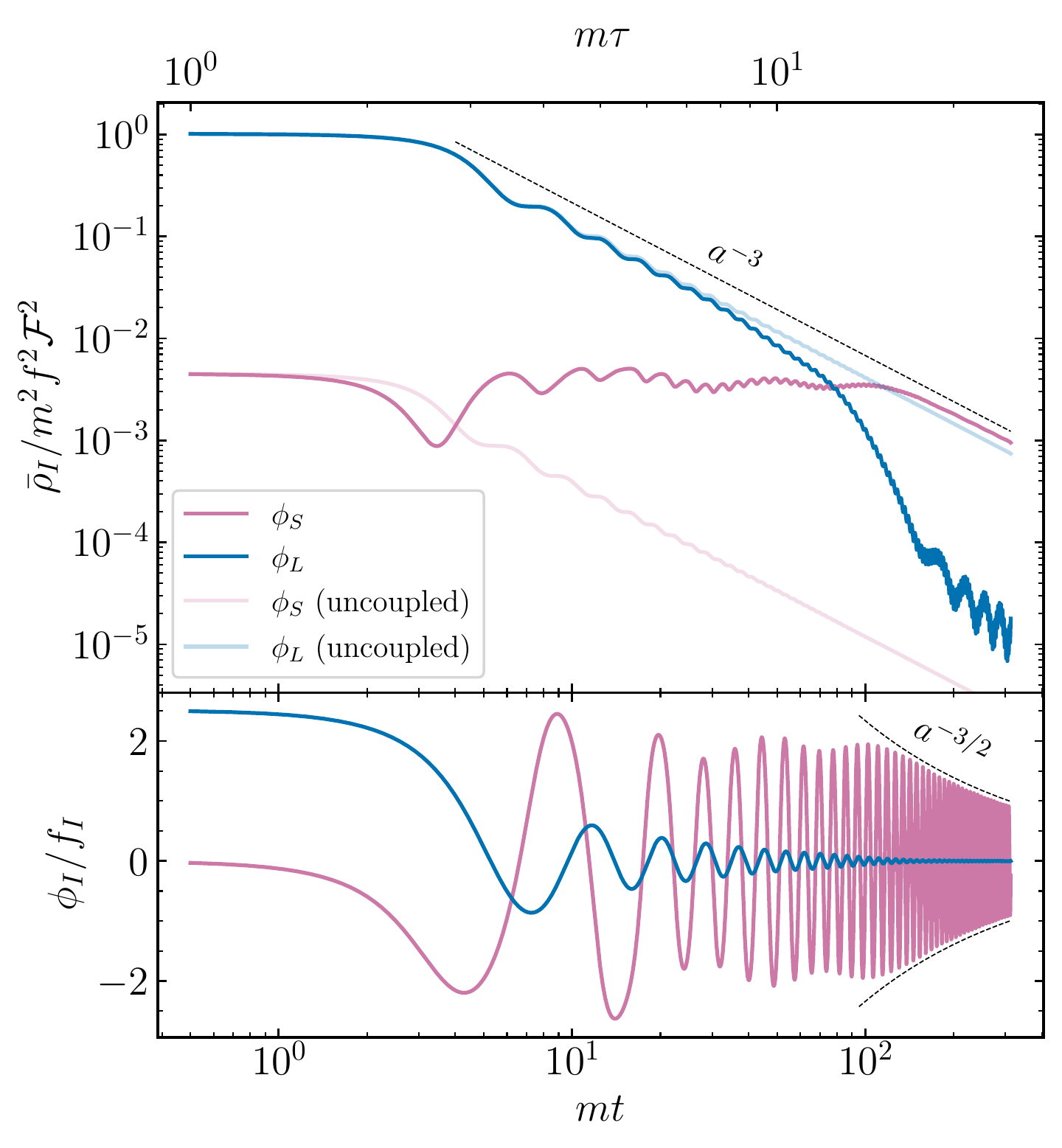}
    \caption{
        Homogeneous dynamics (i.e., ignoring the effect of spatial fluctuations) of friendly axions
        with mass ratio $\mu = 0.75$ and decay constant ratio $\mathcal{F} = 20$.
        Here $t$ and $\tau$ are cosmic and conformal time coordinates, respectively.
        The top panel depicts the evolution of the energy density in the short (pink) and long (blue)
        axions, while the bottom panel displays the field values $\phi_I / f_I$.
        Transparent curves of the same color denote the corresponding results for uncoupled axions
        in the top panel.
    }
    \label{fig:homogeneousExpectation}
\end{figure}

The same conclusion holds for coupled axions as well, so long as the masses of the axions are well
separated.
At large field values, however, interactions can substantially shift the axion oscillation frequency from its ground state value.
Reference~\cite{Cyncynates:2021xzw} showed that coupled axions with a decay constant hierarchy $\calF
\gtrsim 3$ and sufficiently close masses $0.75 \lesssim \mu < 1$ tend to align their frequencies in
a process called autoresonance, illustrated in \cref{fig:homogeneousExpectation}.
Specifically, interactions drive the short axion (with the smaller decay constant) to dynamically
adjust its oscillation amplitude to a fixed value in order to match its frequency to the long axion's,
as evident in the lower panel of \cref{fig:homogeneousExpectation}.
Consequently, the short axion energy density does not dilute like cold matter but instead remains fixed
(as in the top panel of \cref{fig:homogeneousExpectation}) by siphoning energy from the long axion.
If the fields remain spatially homogeneous, this energy transfer runs until backreaction disrupts the precise phase locking of the two fields.
Autoresonance then ends when
$\bar{\rho}_S / \bar{\rho}_L \simeq 2\calF^2(1-\mu)^2 $ for $\calF^2 \gg (1-\mu^2)^{-1}$,
representing a near-complete transfer of the available energy density to the short axion.
In other words, when autoresonance runs to completion, the energy density at late times in the dark
sector is virtually entirely in the short axion---an outcome opposite to what one would
expect from free evolution.

The boost to the late-time energy density of the short axion relative to the scenario of independent axions is of great importance for direct detection experiments.
Laboratory haloscopes probe the couplings of axion DM to SM states, which are typically higher-dimension operators suppressed by the axion decay constant $f_a$.
For example, axions are expected to couple to SM photons via an interaction of the form:
\begin{equation}
    \mathcal{L}
    \supset - \frac{\gagg}{4} \phi F_{\mu \nu} \tilde{F}^{\mu \nu}
\end{equation}
where $\gagg = C_{a \gamma \gamma} \alpha_\text{QED} / 2 \pi f_a$ is the axion-photon
coupling~\cite{Marsh:2015xka,ParticleDataGroup:2020ssz} and $C_{a \gamma \gamma}$ an order-unity
constant.
As discussed in Ref.~\cite{Cyncynates:2021xzw}, when all axions evolve independently from $\mathcal O(1)$ initial misalignment angles, the final energy density $\rho_{a,0}$ of each axion is proportional to $f_a^2$.
In this case, the signal strength $\rho_{a,0}g_{a\gamma\gamma}^2$ is roughly independent of $f_a$;
as such, at a given mass any axion produced by the standard misalignment mechanism would be similarly
hard to see.
In a scenario with friendship however, the boosted late-time energy density of the short axion is $\bar\rho_{S,0} \propto f_L^2$ when autoresonance completes, but the coupling to SM photon is $g_{S\gamma\gamma} \simeq \alpha / 2 \pi f_S$.
Thus the signal strength $\bar \rho_{S,0} g_{S\gamma\gamma}^2$ of the short axion is enhanced by $\calF^2$, making it much more accessible to axion haloscopes.
The effect, however, would be reversed for the long axion: its energy density is suppressed by
$\sim \calF^2$ compared to standard misalignment scenarios.
In this picture, seeing \textit{both} friendly axions would therefore be difficult.

The description of autoresonance given so far assumes the fields remain approximately spatially
homogeneous, but large spatial fluctuations in the axions can prevent the completion of the energy
transfer.
The coherent oscillations of the short axion induce a time-dependent effective mass that
resonantly amplifies fluctuations of the short axion, much like that which characterizes preheating
after inflation~\cite{Traschen:1990sw, Kofman:1994rk, Kofman:1997yn}
(see Refs.~\cite{Bassett:2005xm, Allahverdi:2010xz, Amin:2014eta, Lozanov:2019jxc} for reviews) and large misalignment~\cite{Arvanitaki:2019rax}.
Large-amplitude fluctuations of the short axion can collapse under attractive self-interactions
into oscillons---finite-lifetime, nontopological bound structures with densities of $\mathcal{O}(m^2
f^2)$ and radii of $\mathcal{O}(1/m)$.
Such oscillons explore large field values for the short axion and thus continue to experience large
interactions, but, being nonperturbative objects, are difficult to treat analytically.
Reference~\cite{Cyncynates:2021xzw} presented preliminary evidence that oscillon nucleation occurs for
$\calF \gtrsim 6$ and that oscillons quench autoresonance if they form early enough, setting a limit on the energy density transfer for $\calF\gtrsim 20$.
The remainder of this paper investigates the impact of the nonlinear dynamics of autoresonance and
oscillon formation on the predictions of friendly axion scenarios through the use of $3+1$ dimensional
numerical simulations.

\section{Results} \label{sec:results}

We now present numerical solutions for the fully nonlinear, friendly axion system.
We implement numerical simulations of the axions' classical equations of motions with
\textsf{pystella}~\cite{Adshead:2019lbr,Adshead:2019igv,pystella}, discretizing these equations onto
a $3D$, periodic, regularly spaced grid, computing spatial derivatives via fourth-order centered
differencing, and utilizing a fourth-order Runge-Kutta method for time integration.
Further details are provided in \cref{app:numerical-details}.
Except where otherwise stated, all results use grids with $N^3 = 1024^3$ points, a comoving side
length $L = 1.5 \, \pi / m$ and conformal timestep $\Delta \tau = \Delta x / 10 = L / 10 N$.
The simulations begin with a numerical solution to the linearized system of equations starting
at a time when the Hubble rate $H \ll m$ (see \cref{app:numerical-details} for further details).
The $3D$ evolution begins when $H = m$, corresponding to a conformal time $m \tau_m = 1$
and cosmic time $m t_m = 1 / 2$.
The scale factor is normalized relative to $a_m \equiv a(t_m)$.

Of the free parameters in the model, the decay constant ratio $\calF$ has the strongest effect
on the dynamics.
The mass ratio and initial misalignments mainly determine whether or not autoresonance occurs at
all, whereas the decay constant ratio determines the size of nonlinear backreaction and even whether
fluctuations are sizeably enhanced at all.
Therefore, for most simulations we pick fiducial values $\mu = 0.75$, $\theta_{L}(0, \mathbf{x}) = 0.8 \, \pi$, and
$\theta_{S}(0, \mathbf{x}) = 0$, and run simulations for varying values of $\calF$.\footnote{
    So long as we choose $\theta_L(0, \mathbf{x})$ large enough that the axions experience autoresonance, the initial misalignment angles are essentially inconsequential~\cite{Cyncynates:2021xzw}.
    On the other hand, the choice of the relatively detuned mass ratio $\mu = 0.75$ is made to reduce the runtime of the simulations, as smaller $\mu$ cause perturbations to grow faster (see Appendix~C of Ref.~\cite{Cyncynates:2021xzw}) and shortens the oscillon lifetime
    (explained in \cref{sec:drivenOscillons} below).
}

\begin{figure*}[t!]
    \centering
    \includegraphics[width=\textwidth]{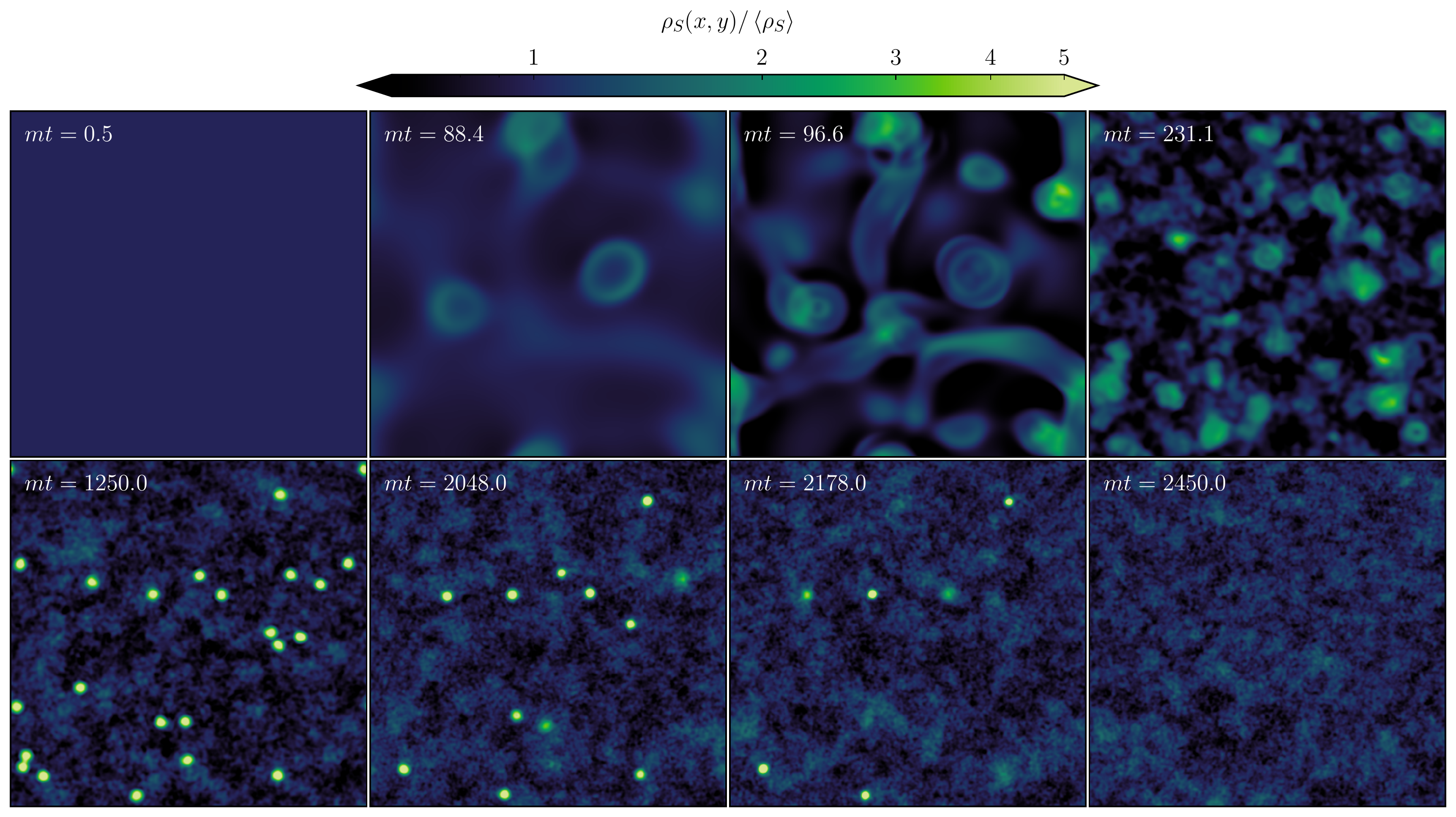}
    \caption{
        Projected density contrast for the short axion, \cref{eqn:def-projected-density-contrast},
        at eight snapshots illustrating key moments during the course of a simulation
        with mass ratio $\mu = 0.8$ and decay constant ratio $\calF = 50$.
        The onset of strong nonlinearity is observed between the third and fourth panel, followed by
        the collapse of large overdensities into spherically symmetric oscillon configurations.
        At this point, autoresonance ceases at the homogeneous level, and energy stops flowing from
        the long axion to the short axion.
        The oscillons persist, driven by the long axion in a form of localized autoresonance
        (see Sec.~\ref{sec:drivenOscillons}).
        Between the sixth, seventh, and eighth panel, the oscillons dissipate, leaving remnant
        overdensities that would eventually seed small-scale galactic substructure.
    }
    \label{fig:delta-S-over-time}
\end{figure*}
For $\calF \lesssim 6$, spatial perturbations do not grow large enough to form oscillons and the
results of the simulations are described completely by Ref.~\cite{Cyncynates:2021xzw}.
For larger $\calF$, fluctuations of $\phi_S$ indeed collapse into oscillons as anticipated by
Ref.~\cite{Cyncynates:2021xzw}.
We present a broad overview of the dynamics of oscillon formation in \cref{fig:delta-S-over-time},
plotting two dimensional projections of the energy density in the short axion at various times over
the course of a simulation.\footnote{
    To be explicit, we display the energy density projected (averaged) along one axis of the
    simulation volume, e.g.,
    \begin{align}\label{eqn:def-projected-density-contrast}
        \rho_{S}(x, y)
        &= \frac{1}{\left\langle \rho_S(x, y, z) \right\rangle}
            \frac{1}{L} \int_{0}^{L} \mathrm{d} z \, \rho_{S}(x, y, z).
    \end{align}
    Such a projected quantity presents more information about the full volume than a single two
    dimensional slice but also underestimates the magnitude of overdensities (since, e.g.,
    any given oscillon occupies only a small fraction of space along the $z$ axis).
}
The field begins in a nearly homogeneous state in the first panel, in which the initial adiabatic
fluctuations are too small to be seen.
The second and third panels depict the linear enhancement of fluctuations by parametric resonance as
the amplification of local overdensities.
Fluctuations become nonlinear at a time $m t_\mathrm{nl} \sim 100$, resulting in large overdensities
that quickly collapse under attractive self-interactions into the oscillons apparent in the fifth
panel.
These oscillons radiate energy and begin to dissipate one by one around $mt \gtrsim 2000$.
Eventually, no bound objects remain and nonlinear interactions cease to dominate the dynamics,
although significant density fluctuations remain.

The interplay between the persistence of homogeneous autoresonance and the onset of nonlinearity
has important consequences for the final distribution of energy between the two axions.
We discuss these dynamics in \cref{sec:energyDensity}, comparing to the results of
Ref.~\cite{Cyncynates:2021xzw}.
In \cref{sec:gravitationalWaves} we compute the gravitational wave production from friendly axions,
finding possible signatures for hyperlight subcomponents in the CMB $B$-mode polarization.
Finally, in \cref{sec:drivenOscillons} we demonstrate that the oscillons that form continue to
experience autoresonance long after the spatially averaged fields cease to resonate, and we discuss the
implications for oscillon lifetimes.

\subsection{Evolution of energy densities}
\label{sec:energyDensity}

\begin{figure}
    \centering
    \includegraphics[width=\columnwidth]{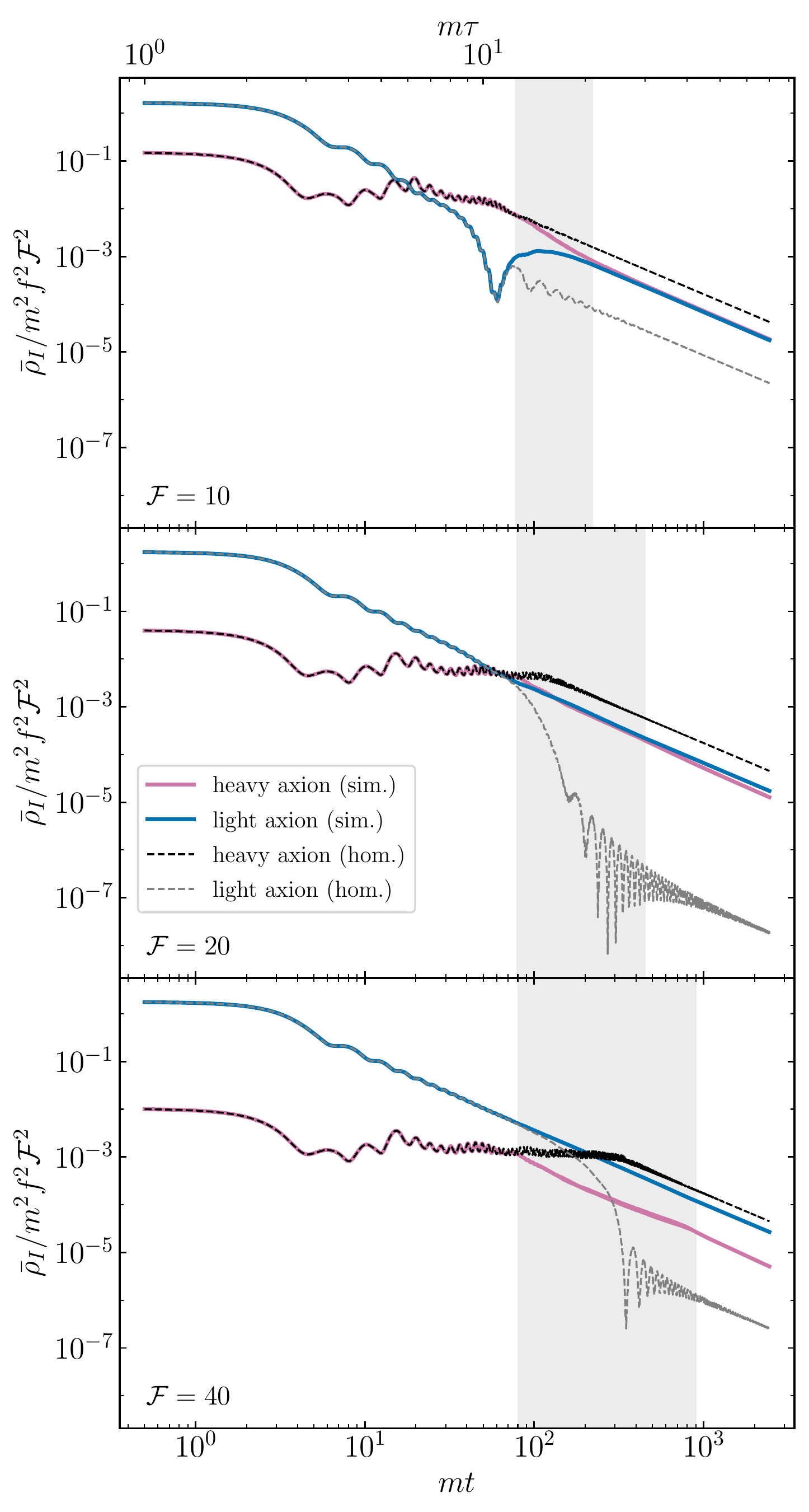}
    \caption{
        Evolution of energy density in the mass eigenstates (\cref{eqn:mass-basis-def}) for three
        simulations with mass ratio $\mu = 0.75$ and decay constant ratios $\calF = 10$, $20$, and $40$ by row.
        Each panel depicts the simulation result for the heavy and light states ($\nu_h$ and $\nu_l$) in pink and blue, respectively, as well as the corresponding results for a purely homogeneous calculation in thin black and gray.
        The shaded region denotes the time when order-one density fluctuations are present, which we
        define as the time when more than 5\% of the energy in the short axion resides in
        overdensities $\rho_S (\mathrm{x}) / \bar{\rho}_S > 10$.
        Shortly after these large overdensities form, they either dissipate or coalesce into oscillons, so the gray regions are decent proxies for the presence of oscillons.
        Note that in contrast to \cref{fig:homogeneousExpectation}
        we here plot the energy density of the mass-basis fields rather than the interaction-basis
        ones.
    }
    \label{fig:rho-evolution}
\end{figure}

Having established the importance of nonlinear dynamics for a large portion of parameter space, we
now investigate how nonlinear density fluctuations impact the final distribution of energy between
the two axions (and, as a consequence, their relic abundances today).
We first study the evolution of each axion's energy density in \cref{fig:rho-evolution} for three
representative values of $\calF$, comparing the result of simulations to that of a homogeneous
analysis.
To avoid ambiguities in the final partition of energy densities we work in the mass basis
\begin{subequations}\label{eqn:mass-basis-def}
\begin{align}
    \nu_h
    &\equiv \phi_S\cos\eta + \phi_L\sin\eta, \\
    \nu_l
    &\equiv -\phi_S\sin\eta + \phi_L\cos\eta, \\
    \cos2\eta
    &\equiv \frac{1 - \mu^2-\calF^{-2}}{\sqrt{4\calF^{-2} + (1 - \mu^2 - \calF^{-2})^2}},
\end{align}
\end{subequations}
where the heavy state $\nu_h$ is composed mostly of the short axion, and the light state $\nu_l$ is composed mostly of the long axion in the limit $\calF\gg 1$ (see Appendix~A of Ref.~\cite{Cyncynates:2021xzw} for a complete discussion).
Each panel exhibits an initial phase of homogeneous, autoresonant energy transfer and the onset of
nonlinearity that quenches autoresonance, at which point the energy density departs from the trend
of the homogeneous result.
From analytic estimates of growth rate for the fastest growing mode~\cite{Cyncynates:2021xzw}, nonlinearity occurs at approximately\footnote{This result accounts for both Hubble friction and the slight decay of the initial metric perturbation before the fastest-growing mode starts growing (see Eqs.~C17 and C18 in Ref.~\cite{Cyncynates:2021xzw} and the surrounding discussion, fixing $\delta\omega = \mu - 1$).}
\begin{align}\label{eqn:tnl-approximation}
    m t_\mathrm{nl}
    &\approx 17.6 \frac{1 - 0.1 \log(1 - \mu)}{1 - \mu},
\end{align}
in good agreement with $m t_\mathrm{nl} \approx 80$ observed in \cref{fig:rho-evolution}.
The ultimate partitioning of energy depends primarily on the precise timing of nonlinearity and
oscillon formation relative to the (would-be) completion of autoresonance, a point which we detail
below.
We now describe these two regimes of $\calF$ in detail.

For $6 \lesssim \calF \lesssim 20$, oscillons nucleate \textit{after} the short axion's energy
density first exceeds the long axion's.
At roughly the same time, autoresonance ends and $\bar{\rho}_h$ ceases to be roughly constant, instead
decaying approximately as $a^{-3}$ like nonrelativistic matter.
Contrary to the homogeneous analysis of Ref.~\cite{Cyncynates:2021xzw}, however, we observe in this
range that nonperturbative dynamics in fact enable energy transfer from the short axion back to the long axion,
resulting in late-time \mbox{(near-)equilibration} of the two axion energy densities.
This phenomenon is most evident in the top panel of \cref{fig:rho-evolution} ($\calF = 10$),
where the heavy and light axions' energy densities asymptote toward a common value.

Interactions between the two axions are strongest where the field values are largest, suggesting
that oscillons play a key role in reversing energy transfer.
Inside an oscillon, the field amplitude oscillates with a period $\omega < m$ due to its binding
energy.
Since the long axion's natural frequency is $\mu m < m$, an oscillon can provide a locus for more
efficient energy transfer from the short axion back to the long axion.\footnote{
    In fact, during autoresonance the short axion is driven at exactly the frequency $\mu m$.
    When fluctuations grow nonperturbative the oscillon frequencies will thus remain close to $\mu m$.
}
Indeed, for most decay constant ratios $6 \lesssim \mathcal{F} \lesssim 20$, the end of autoresonance
and formation of oscillons is associated with a substantial transfer of energy to the light axion.
For $6 \lesssim \calF \lesssim 10$, the final stage of energy transfer to the light axion occurs
in discrete jumps that appear to coincide with the death of individual oscillons.
In all cases, we observe that most of the radiation from the heavy axion into the light axion
is into semirelativistic modes, as one would expect if oscillons are responsible for equilibration.
However, at larger $\calF$ equipartition is nearly achieved by the time oscillons form anyway;
the subsequent evolution is more continuous, obfuscating any association between oscillon death
and energy transfer.
While nonlinear effects are evidently crucial, identifying the specific mechanism for energy flow in
general is challenging and would require software infrastructure and computational resources---in
particular, substantial storage of $3D$ snapshots and analysis thereof---well beyond the reach of
the present work.

The middle panel with $\calF = 20$ represents the marginal case where oscillons form at nearly
the exact time that the heavy axion's energy density first reaches that of the light axion.
For larger values $\calF \gtrsim 20$, $t_\mathrm{nl}$ and oscillon formation occur before the heavy
axion dominates the sector's energy density, terminating autoresonant energy transfer to the heavy
axion.
As shown in the bottom panel of \cref{fig:rho-evolution} ($\calF = 40$), the energy density in both
axions then decays as approximately $a^{-3}$.
In this case the backreaction effects at play for smaller $\calF$ are too suppressed to enable
substantial energy transfer by the oscillons.
The trends for yet larger $\calF$ are qualitatively similar: parametric resonance proceeds at the
same rate and oscillons form at a similar time.
The final ratio of energy densities $\bar{\rho}_h / \bar{\rho}_l$ thus receives a constant boost due
to the period of autoresonance but still decreases as $1 / \calF^2$.

\begin{figure}[t!]
    \centering
    \includegraphics[width=\columnwidth]{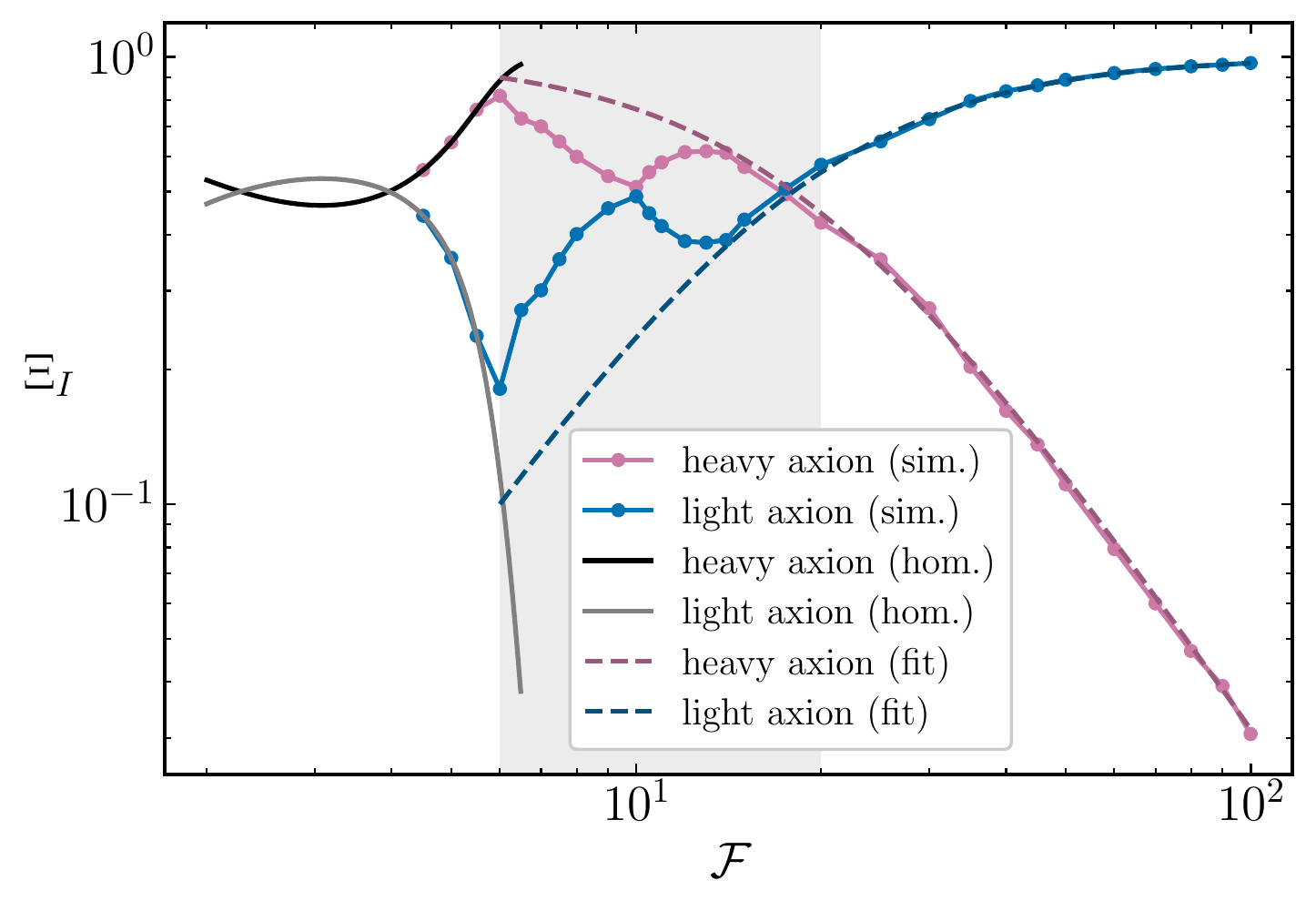}
    \caption{
        Late-time partition of energy between the friendly axions
        ($\Xi_I$, \cref{eqn:energy-partition-def}) as a function of the decay constant ratio $\calF$,
        evaluated at the end of the simulations (long after energy transfer ceases).
        The mass ratio is fixed to $\mu = 0.75$.
        Results for the heavy and light axions are respectively in pink and blue for the simulations
        and black and gray for the homogeneous computation.
        The dashed, dark pink and blue curves depict the empirical fit of
        \cref{eqn:energy-partition-estimate}.
        The shaded region indicates the range of $\calF$ for which oscillons form
        after the heavy axion's energy overtakes the light one's (in which case nonlinear effects
        return a non-negligible amount of energy to the light axion).
    }
    \label{fig:final-rho-vs-F}
\end{figure}

Having discussed the dynamics that control the distribution of energy between the two axions,
we now summarize the full $\calF$-dependence of the late-time energy fractions,
\begin{align}\label{eqn:energy-partition-def}
    \Xi_{I}
    &\equiv \left. \frac{\bar{\rho}_I}{\bar{\rho}_h + \bar{\rho}_l} \right\vert_\text{late time},
\end{align}
where $I = h, l$. The final partitioning changes qualitatively at a critical decay constant ratio $\calF_\star$
for which nonlinearities become important (at $t_\mathrm{nl}$) just as the heavy axion's energy
density first matches the light one's (via autoresonant energy transfer).
From our simulations we find $\calF_\star \approx 20$ for $\mu = 0.75$; this value depends on
the mass ratio in the same manner as $t_\mathrm{nl}$ (c.f. \cref{eqn:tnl-approximation}).
This timing separates two distinct regimes: one of near-equilibration due to nonlinear effects
at $\calF < \calF_\star$ and a $1/\calF^2$ suppression of the heavy-axion abundance via the early end of
autoresonance at larger $\calF$.
Both regimes are well captured by
\begin{subequations}\label{eqn:energy-partition-estimate}
\begin{align}
    \label{eq:frach}
    \Xi_h
    &\sim
        \begin{dcases}
            \frac{1}{2}
            & 6 \lesssim \calF \lesssim \calF_\star \\
            \frac{1}{1 + 1.3 (\mathcal{F} / \mathcal{F}_\star)^2}
            \hphantom{1 - }
            & \calF \gtrsim \calF_\star
        \end{dcases} \\
    \label{eq:fracl}
    \Xi_l
    &\sim
        \begin{dcases}
            \frac{1}{2}
            & 6 \lesssim \calF \lesssim \calF_\star \\
            1 - \frac{1}{1 + 1.3 (\mathcal{F} / \mathcal{F}_\star)^2}
            & \calF \gtrsim \calF_\star
        \end{dcases},
\end{align}
\end{subequations}
including an empirical factor of $1.3$ that best fits the results from simulations.
We display the corresponding quantities computed directly from simulations in
\cref{fig:final-rho-vs-F}.
For $\mathcal{F} \lesssim 6$, the energy densities indeed match those predicted by the
homogeneous theory, which are computed in full in Ref.~\cite{Cyncynates:2021xzw}.
For such low decay constant ratios, autoresonance is too brief to support parametric resonance long
enough for perturbations to reach order unity.
For $6 \lesssim \calF \lesssim 20$, the energy density of the light axion, instead of being entirely
depleted, remains within a factor of between 1 and 4 of the heavy axion's energy density with a
precise dependence on $\calF$ beyond the sophistication of \cref{eqn:energy-partition-estimate}.
At larger $\calF \gtrsim 20$, the $1 / \calF^2$ scaling takes over, parametrically suppressing
the heavy axion's abundance relative to the light axion's.
Nonetheless, in this range the heavy axion carries approximately $\calF_\star^2/1.3 \sim 310$ times
more energy density than it would have had in the absence of autoresonance.

While the above results fix the mass ratio to $\mu = 0.75$, we do not expect our findings to differ
substantially in the parameter space where autoresonance occurs.
However, we do expect the precise thresholds in $\calF$ of each regime to vary mildly with $\mu$.
In particular, $\calF_\star$ varies in accordance with the change in the timescale of
nonlinearity $t_\mathrm{nl}$, which itself varies with $\mu$ via \cref{eqn:tnl-approximation}.
Limited simulations corroborate these expectations for $0.70 \lesssim \mu \lesssim 0.80$, the range
for which simulation runtimes are tractable, but a more quantitative assessment of the variation of
trends with $\mu$ would require multiple times more computational resources.
For increasingly degenerate axion masses, linear resonance becomes more relevant and could
qualitatively alter our findings for $0.95 \lesssim \mu \leq 1$.

The light axion's enhanced abundance relative to homogeneous results has important consequences for
direct detection experiments.
Although the near-even partitioning of energy for $6 \lesssim \calF \lesssim \calF_\star$ implies
that the heavy axion is slightly harder to detect than predicted by Ref.~\cite{Cyncynates:2021xzw},
it also implies that the light axion requires only twice the experimental sensitivity as it would for
an uncoupled, misaligned axion of the same mass.
The homogeneous expectation, in contrast, was that the light axion would require $\calF^2$ times
greater experimental sensitivity to detect.
The nonperturbative equalization of energy density in the sector thus serves to make the sector
\textit{as a whole} more visible to direct detection experiments.
Observing two axions with similar masses---with the heavier axion substantially more visible than
expectations for single-axion misalignment, and the lighter one comparably so---is a characteristic
signature of friendly dynamics of the form described here.
For the mass range $10^{-10} \eV \lesssim m \lesssim 10^{-3} \eV$, many near-future experimental
efforts will probe relevant parameter space (see, e.g.,
Refs.~\cite{Brouwer:2022bwo,Alesini:2017ifp,Stern:2016bbw,BRASS,Lasenby:2019prg,Berlin:2019ahk,Berlin:2020vrk,DMRadio:2022pkf,Beurthey:2020yuq,McAllister:2017lkb,Nagano:2019rbw,Liu:2018icu}).

To close, we connect the partitioning of \cref{eqn:energy-partition-estimate} to present-day
abundances by estimating the net present-day energy density in the sector.
The energy density at horizon crossing is dominated by the light axion, i.e., $\rhotot ( t_m ) \sim
\mu^2 m^2 \calF^2 f^2 \Theta_{L,0}^2$, which subsequently redshifts as $a^{-3}$.\footnote{
    The $a^{-3}$ redshifting assumes that the axions are always noninteracting and nonrelativistic.
    In reality, at early times the axion interactions during autoresonance cause $\rhotot$ to redshift slightly slower than $a^{-3}$, and at later times oscillons radiate mildly relativistic axions such that $\rhotot$ redshifts slightly faster
    than $a^{-3}$ (until all axions become nonrelativistic).
    Together, these effects amount to only an $\mathcal{O}(1)$ factor which we neglect for simplicity.
}
Combined with \cref{eqn:energy-partition-estimate}, the present-day abundance of the each axion is
\begin{align}\label{eqn:final-relic-abundance}
\begin{split}
    \frac{\Omega_{h, 0}}{0.13}
    &\approx
        \Xi_I
        \left(
            \frac{m}{10^{-19} \, \mathrm{eV}}
        \right)^{1/2}
        \left(
            \frac{\mathcal{F} f}{10^{16} \, \mathrm{GeV}}
        \right)^2
        \left( \mu \Theta_{L, 0} \right)^2,
\end{split}
\end{align}
with $\Xi_I$ set by \cref{eqn:energy-partition-estimate}.
Factors accounting for the thermal history of the SM (i.e., the number of effective relativistic
degrees of freedom in the SM entropy and energy density) change the above result by only an
$\mathcal{O}(1)$ factor over the mass range of interest and are omitted for simplicity.

\subsection{Gravitational waves}
\label{sec:gravitationalWaves}

Rapidly growing fluctuations during and after autoresonance and the resulting oscillons can
both source gravitational waves, again much like the parametric resonance and oscillon formation
that can occur during preheating~\cite{Khlebnikov:1997di, Easther:2006gt, Easther:2006vd,
Easther:2007vj, Garcia-Bellido:2007nns, Dufaux:2007pt, Zhou:2013tsa, Lozanov:2019ylm, Amin:2018xfe,
Hiramatsu:2020obh} and single-axion misalignment~\cite{Arvanitaki:2019rax}.
In this section we compute the signal strength generated by friendly axions and discuss the
corresponding constraining power of existing and future observations.
\Cref{app:gravitational-waves} briefly reviews stochastic gravitational wave backgrounds and
the transfer functions required to relate their spectra at emission to that at the present day.

We begin by estimating the scaling of the peak amplitude of the gravitational wave spectrum with the
short-axion mass $m$ and long axion's decay constant $\mathcal{F} f$.
Gravitational waves are sourced by the anisotropic part of the stress tensor (via
\cref{eqn:gw-eom}), whose components scale like the energy density of the source.
The time (relative to $t_m$) and wave number (relative to $m$) of peak emission varies weakly with
model parameters (and is entirely independent of $f$).
The spectral abundance of gravitational waves (\cref{eqn:omega-gw-spectrum-ito-power-spectrum})
therefore scales with two powers of the fractional energy density of the source at the time of
emission, $\bar{\rho}_\mathrm{source}(t) / \bar{\rho}(t)$.
The short axion is the dominant source of anisotropic stress, which we expect to peak near the time
when the system becomes nonlinear, $t_\mathrm{nl}$, when axion gradients are largest and power is
scattered to smaller scales.

To proceed, we follow the model-independent heuristics of Refs.~\cite{Giblin:2014gra, Amin:2014eta}.
Approximating the gravitational wave source as a Gaussian peaking at
momentum $k_\star$ with width $\sigma$, one may estimate the peak amplitude to
be~\cite{Giblin:2014gra}:
\begin{align}\label{eqn:rule-of-thumb-1}
    \Omega_{\mathrm{GW}}(k_\star)
    &= \frac{27 \gamma^2 \nu^2}{\sqrt{\pi}}
        \frac{k_\star}{\sigma}
        \left( \frac{a H_p}{k_\star} \right)^2,
\end{align}
at the time the source is maximized.
Here $\gamma$ is what fraction of the Universe's energy is in the source at the time of the process,
$\nu$ measures how anisotropic the source is, and $H_p$ the Hubble parameter at the time of the
process.\footnote{
    Note that our $\nu^2$ corresponds to $\beta w^2$ in terms of the parameters of
    Ref.~\cite{Giblin:2014gra}.
}
The peak wave number $k_\star$ and width $\sigma$ are straightforward to approximate (or read off of
simulation results), but the anisotropy coefficient $\nu$ is harder to estimate;
Ref.~\cite{Giblin:2014gra} motivates $\nu \sim 10^{-2}$ to $10^{-1}$ for typical processes.
Evaluating \cref{eqn:rule-of-thumb-1} at $t_\mathrm{nl}$ and plugging in
$a H = 1 / \sqrt{2 t_\mathrm{nl}}$,
\begin{align}\label{eqn:rule-of-thumb-2}
    \Omega_{\mathrm{GW}}(k_\star)
    &\approx
        \frac{27 \nu^2}{\sqrt{\pi}}
        \frac{k_\star}{\sigma}
        \left( \frac{m}{k_\star} \right)^{2}
        \frac{\left( \mu \mathcal{F} \Theta_{L, 0} f / \Mpl \right)^4
        }{
            \left[ 1 + 1.3 \left( \mathcal{F} / \mathcal{F}_\star \right)^2 \right]^2
        },
\end{align}
where we have used that the energy density in the short axion at $\tnl$ is approximately
$1 / [1 + 1.3 (\calF / \calF_\star)^2]$
of the total axion energy density, and the total axion energy density is
given by $\rhotot(t_m) (t_m / \tnl )^{3/2}$.
Notice that the suppression from how far inside the horizon the peak is (the factor of
$[a H_p / k_\star]^2$ in \cref{eqn:rule-of-thumb-1}) is exactly compensated by the growth in time
of $\gamma$ (since the homogeneous energy available to source the short axion redshifts with one
fewer power of the scale factor than the SM radiation).
We therefore expect gravitational wave signals from friendly axions to be only weakly sensitive to
the time of nonlinearity $t_\mathrm{nl}$.

By comparing \cref{eqn:rule-of-thumb-2} with the relic abundance (\cref{eqn:final-relic-abundance})
we may estimate the peak of the gravitational wave signal as a function of the mass $m$
and relic abundance of the heavy axion $\Omega_h(t_0)$:
\begin{align}
\begin{split}
    \Omega_{\mathrm{GW}, 0} h^2
    &=
        \frac{27 \nu^2}{\sqrt{\pi}}
        \frac{k_\star}{\sigma}
        \left( \frac{m}{k_\star} \right)^{2}
        \left[ \Omega_{\mathrm{rad}}(t_0) h^2 \right]^{-1/2}
    \\ &\hphantom{ {}={} }
        \times
        \left[ \Omega_h(t_0) h^2 \right]^2
        \left(
            \frac{m}{H_{100}}
        \right)^{-1}
\end{split}
\end{align}
where we have dropped factors of the relativistic degrees of freedom, which reduce the amplitude by
at most a factor of two at early enough times $t_m$ such that all SM species are in thermal equilibrium.
Here, $H_{100} = 100 \, \mathrm{km} / \mathrm{s} / \mathrm{Mpc} = 2.13 \times 10^{-33} \, \mathrm{eV}$ and $h = H_0/H_{100}$.
From our simulations we observe that $k_\star / m \approx 9$ and $\sigma \approx k_\star / 3$.
Taking $\nu = 1/20$ (for which the estimates agree well with the simulation results)
and considering
the regime $\mathcal{F} \lesssim 20$ for which the signal is not suppressed, the peak amplitude is
\begin{align}\label{eqn:omega-gw-scaling}
    \Omega_{\mathrm{GW}, 0}(k_\star) h^2
    &\approx 10^{-15}
        \left(
            \frac{\Omega_S(t_0) h^2}{0.06}
        \right)^{2}
        \left(
            \frac{m}{10^{-21} \, \mathrm{eV}}
        \right)^{-1},
\end{align}
at a present-day frequency of the peak of
\begin{align}\label{eqn:gw-frequency-scaling}
    \frac{f_{\mathrm{GW}, \star}}{\mathrm{Hz}}
    &\approx 2.8 \times 10^{-14}
        \frac{k_\star}{m}
        \left(
            \frac{m}{10^{-21} \, \mathrm{eV}}
        \right)^{1/2}.
\end{align}
The amplitude estimate \cref{eqn:omega-gw-scaling} agrees quite well---within a factor of a few---with
the spectra from simulations when evaluated at $t_\mathrm{nl}$.
However, the spectra evaluated at the end of the simulation (after all gravitational wave production has concluded) are about an order of magnitude larger due to factors not captured by
these simplistic estimates such as the time evolution of the source after $t_\mathrm{nl}$.

From \cref{eqn:omega-gw-scaling,eqn:gw-frequency-scaling} we see that adjusting the mass
$m$ to change the peak frequency by some factor modulates the gravitational wave power spectrum
by two powers of that factor.
Signals that could be visible at pulsar timing arrays (with frequencies of order
$10^{-9} \, \mathrm{Hz}$) could therefore not exceed amplitudes of $10^{-23}$
(about eight orders of magnitude below the projected sensitivity of the Square Kilometer Array~\cite{Janssen:2014dka,Schmitz:2020syl})
without requiring an axion abundance that would overclose the Universe.
This is an unfortunate consequence of the source both being short-lived and
redshifting more slowly than the SM plasma from the time of gravitational wave production
to the present day.\footnote{
    Despite the dearth of direct gravitational-wave probes at frequencies between $10^{-15}$
    and $10^{-9}$ Hz, two \textit{indirect} constraints apply around this range.
    Precision CMB measurements of the energy density in radiation in the Universe provide an upper
    bound on the present-day gravitational wave spectrum (gravitational waves themselves contributing to
    expansion like radiation)~\cite{Maggiore:1999vm, Dvorkin:2022jyg}, currently of order
    $\Omega_{\mathrm{GW}, 0} h^2 \sim 10^{-6}$~\cite{Planck:2018vyg,Pagano:2015hma}.
    In addition, recent work has argued that spectral distortions of the CMB blackbody also probe
    gravitational waves in this frequency range~\cite{Kite:2020uix}.
    While far-future experiments could provide tighter constraints than
    $N_\mathrm{eff}$ measurements, these still are unlikely to be useful probes of friendly axions.
}
Note that the stochastic background from single-field inflation, if detectable by future CMB
experiments, is nearly scale invariant and of order $10^{-16}$~\cite{Smith:2005mm,Caprini:2018mtu},
far larger than those possible from friendly axion DM.

On the other hand, existing measurements of the $B$-mode polarization of the CMB already constrain
gravitational waves at frequencies between $10^{-18}$ and $10^{-16}$ Hz, with a most stringent upper
limit of $\Omega_{\mathrm{GW}, 0} h^2 \sim 10^{-16}$ at
$f_\mathrm{GW} \sim 10^{-17} \, \mathrm{Hz}$~\cite{Clarke:2020bil}.
Importantly, the polarization is sourced almost exclusively at recombination, when the photon visibility
function spikes.
As a result, CMB constraints are only relevant for scenarios where gravitational waves are sourced
before this time, i.e., when the Hubble scale is
$H \gtrsim H_\mathrm{rec} \sim 3 \times 10^{-29} \, \mathrm{eV}$~\cite{Planck:2018vyg}.
In the scenarios we consider here, the anisotropic stress maximizes after about ten field oscillations,
so the smallest relevant mass is of order $10^{-27} \, \mathrm{eV}$.
Consulting \cref{eqn:gw-frequency-scaling}, the peak of the signal at such a mass corresponds to a
frequency of order $2 \times 10^{-16} \, \mathrm{Hz}$.
Such hyperlight axion dark matter is well ruled out by fuzzy dark matter constraints,
but could make up some subcomponent of the total dark matter (depending on the mass)~\cite{Irsic:2017yje,Armengaud:2017nkf,Dentler:2021zij,Bozek:2014uqa,Hlozek:2014lca,Kobayashi:2017jcf,DES:2018zzu,Lague:2021frh,Hlozek:2017zzf,Dalal:2022rmp,Flitter:2022pzf}.

By the preceding argument, CMB measurements mainly probe the infrared tail of the gravitational wave
background from friendly axions.
Simulating a large enough volume to capture these modes while also resolving the nonlinear dynamics at
small scales requires orders of magnitude more computational resources than available to us.
Fortunately, on causal grounds the gravitational
wave spectrum on scales much larger than the relevant dynamical scales (i.e., $k / a \ll m$) follows a
simple power law, independent of the underlying dynamics~\cite{Hook:2020phx}.
We therefore extrapolate the signals computed in simulations as decaying with $f_\mathrm{GW}^3$ at smaller
frequencies, appropriate for infrared, ``causal'' modes generated inside the horizon and those generated
while superhorizon that reenter during radiation-dominated era.
The lowest-frequency modes in the simulation do nearly follow an $f_\mathrm{GW}^3$ scaling, so we expect
this approximation to be at worst conservative.
Constraints were similarly derived on the infrared tail of gravitational waves from a model
of early dark energy in Ref.~\cite{Weiner:2020sxn}.
However, note that recombination occurs shortly after matter-radiation equality, and superhorizon
causal  modes that reenter the horizon during the matter era instead grow with one
power of $f_\mathrm{GW}$.\footnote{
    See Ref.~\cite{Hook:2020phx} for a thorough presentation of the imprint of the
    expansion history and presence of free-streaming radiation on causal gravitational waves.
}
Since the CMB becomes increasingly less sensitive to $\Omega_\mathrm{GW}$ on scales larger than
the horizon at equality, we do not expect the break in the power law to improve constraints.
(On the other hand, it would likely be observable for causal gravitational wave backgrounds
that are indeed observable in the CMB.)
Our simulations themselves also do not account for the transition to matter domination,
instead assuming a radiation-dominated Universe;
we expect this to be entirely sufficient for our estimates here.

\begin{figure}[t!]
    \centering
    \includegraphics[width=\columnwidth]{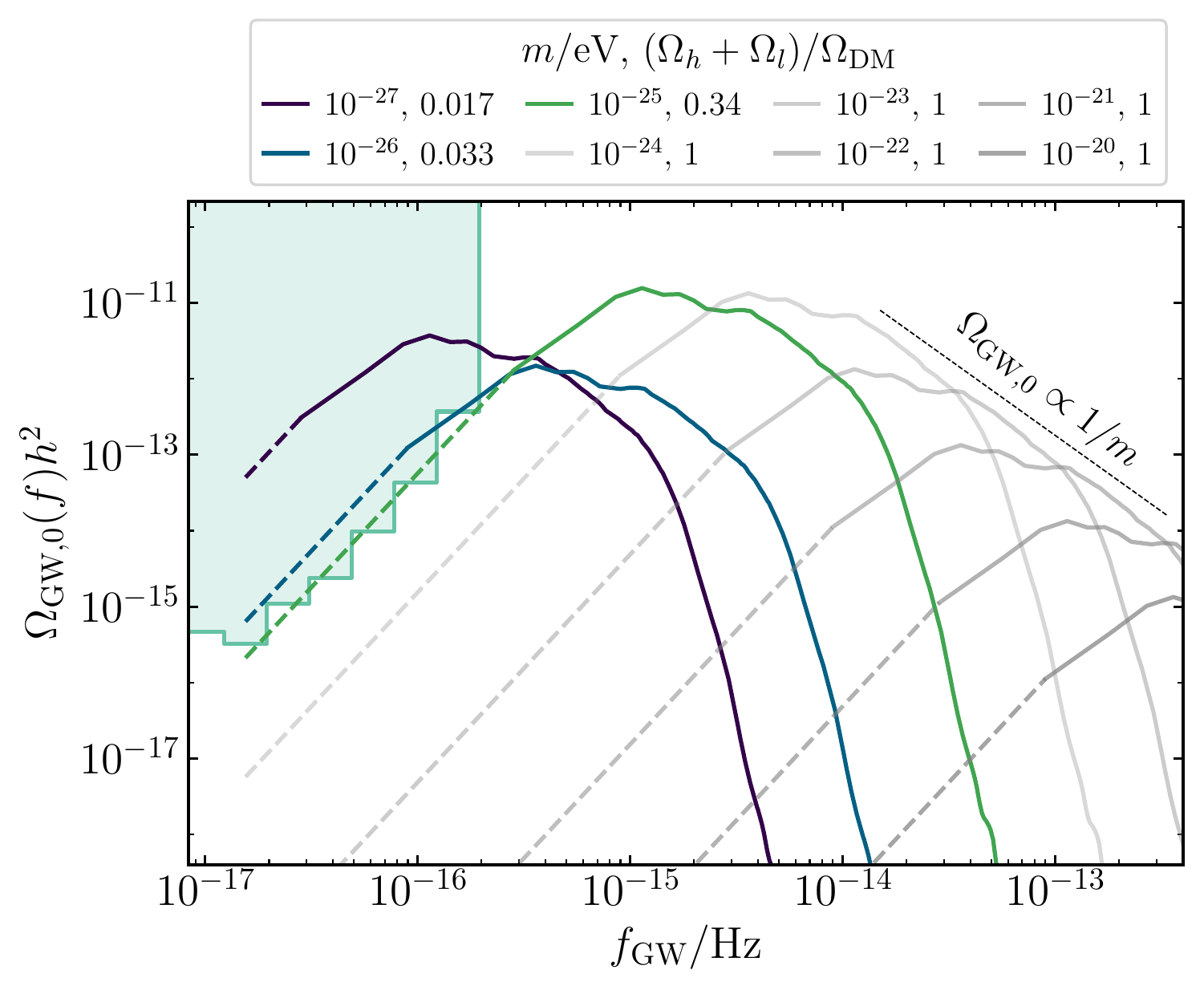}
    \caption{
        Gravitational wave backgrounds from friendly axions
        with fixed decay constant ratio $\mathcal{F} = 20$ and mass ratio $\mu = 0.75$.
        Each curve depicts the signal for a short axion mass making up a fraction of the
        total dark matter $F_\mathrm{DM}$ as allowed by CMB and large scale structure
        data~\cite{Lague:2021frh}, indicated by the legend.
        Dashed lines indicate the extrapolation of the signal computed in the simulations
        (solid lines) to smaller frequencies as an $f_\mathrm{GW}^3$ power law as justified in the text.
        The transparent gray curves, with masses for which the aforementioned probes do not provide
        constraints, take the axions to make up all of dark matter.
        Such light dark matter is ruled out by fuzzy dark matter constraints, but
        we include these curves to illustrate the mass dependence of the signals.
        CMB constraints from Ref.~\cite{Clarke:2020bil} are superimposed in light green.
    }
    \label{fig:gws-at-cmb}
\end{figure}

We investigate the possibility of CMB constraints in \cref{fig:gws-at-cmb}, which displays
the possible signals from friendly axions subcomponents with $\calF = 20$.
To illustrate the constraining power on the present abundance of friendly axions,
we consider masses varying from $10^{-27}$ to $10^{-25} \, \mathrm{eV}$, each
taking the fraction of dark matter in friendly axions to saturate limits on ultralight axions
from CMB and large scale structure data~\cite{Lague:2021frh}.
Because the CMB is highly sensitive to low-frequency tensor modes (in terms of their effective
energy density)~\cite{Clarke:2020bil}, the CMB can place tight constraints on the fraction
of dark matter in hyperlight friendly axions.
For $m = 10^{-27} \, \mathrm{eV}$, Planck and BICEP2/\textit{Keck}~\cite{Planck:2018jri,BICEP2:2018kqh}
allow only a $0.1 \%$ friendly subcomponent.\footnote{
    Note that the most recent BICEP/\textit{Keck} data release further improved constraints
    on the tensor-to-scalar ratio by a factor of $\sim 1.6$, and that CMB-S4 projects to
    provide upwards of a further factor of $\sim 30$~\cite{CMB-S4:2020lpa}.
}
We may obtain a heuristic bound as a function of mass by extrapolating the $f_\mathrm{GW}^3$
tail to the frequency corresponding to the horizon size at matter-radiation equality,
$f_\mathrm{eq} = 1.54 \times 10^{-15} \, \mathrm{Hz}$, where Ref.~\cite{Clarke:2020bil} reports
an upper limit of $\Omega_{\mathrm{GW}, 0}^\mathrm{bound} h^2 = 3.2 \times 10^{-16}$.
Applying the scalings of \cref{eqn:omega-gw-scaling,eqn:gw-frequency-scaling}, the CMB probes
\begin{align}
    \frac{\Omega_h + \Omega_l}{\Omega_\mathrm{DM}}
    \lesssim 10^{-2}
        \left( \frac{m}{5 \times 10^{-27} \, \mathrm{eV}} \right)^{5/4}
        \sqrt{
            \frac{\Omega_{\mathrm{GW}, 0}^\mathrm{bound}(f_\mathrm{eq}) h^2}{ 3.2 \times 10^{-16} }
        }
\end{align}
which is only relevant for masses $m \gtrsim 10^{-27} \, \mathrm{eV}$ for which the signals are
produced before recombination.
With the bound of Ref.~\cite{Clarke:2020bil}, the limits become irrelevant (i.e., order unity)
for masses $m \gtrsim 10^{-25} \, \mathrm{eV}$, well below the present lower limits on the mass
of fuzzy dark matter.

Finally, for illustrative purposes \cref{fig:gws-at-cmb} includes signals for friendly axions with
larger masses $m > 10^{-25} \, \mathrm{eV}$ taken to make up all of the dark matter.
While these scenarios themselves are ruled out by lower limits on the mass of axion dark matter,
they demonstrate that gravitational waves from friendly axions would likely be unobservably small at
allowed masses.
Even for $m = 10^{-20} \, \mathrm{eV}$ the peak of the signal is only $\sim 10^{-15}$, consistent
with the preceding discussion.

\subsection{Driven oscillons}
\label{sec:drivenOscillons}

Beyond their important role in the nonperturbative dynamics of friendly axions, oscillons have long been a subject of interest~\cite{kudryavtsev1975solitonlike,Makhankov:1978rg,Kolb:1993hw,Salmi:2012ta,Amin:2011hj,Kawasaki:2019czd}.
As nontopological excitations, oscillons generically decay by radiating semirelativistic modes, making their lifetime an interesting object of study~\cite{Olle:2020qqy,Zhang:2020bec,Cyncynates:2021rtf}.
In models with friendly axions, the lifetime of an oscillon can be extended beyond its in-vacuum expectation because of energy transfer from the long axion~\cite{Cyncynates:2021xzw}.
In this section, we show that short axion oscillons are driven by the long axion via autoresonance, obeying equations analogous to those for the homogeneous fields.\footnote{
    Other related nonlinear wave equations exhibit oscillons/solitons driven by autoresonance, e.g., Refs.~\cite{friedland2005excitation,friedland2006emergence,maslov2007breather,batalov2011control,karachalios2019excitation}.
}
We then provide analytic results to quantify the oscillon lifetime enhancement and numerical results to support our analysis.
Because of the importance of nonlinear dynamics, it is more convenient to discuss oscillons in the interaction basis of the short and long axion, which we adopt consistently throughout this section.

An oscillon is a quasiperiodic, quasilocalized excitation of the axion field, and its
fundamental frequency $\omega$ is smaller than the rest mass of the axion $m$.
After its initial formation, the binding energy per particle inside an isolated oscillon,
\begin{align}
    E_\te{bind} = m - \omega,
\end{align}
is a decreasing function of time, and consequently the characteristic size of the oscillon
\begin{align}
    r_\te{osc} \approx \f{\pi}{\sqrt{2 m E_\te{bind}}},
\end{align}
is an increasing function of time.
Over time its rate of radiation falls off approximately exponentially with this growing separation of scales~\cite{Cyncynates:2021rtf}:
\begin{align}
    P_\te{rad}/f^2\sim \exp\ps{-\f{r_\te{osc}}{\lambda_\te{rad}}}\sim \exp\ps{-\f{3\omega}{\sqrt{2 m (m - \omega)}}},
\end{align}
where $\lambda_\mathrm{rad} = 3\omega$ is the radiation wavelength due to $3\to1$ processes, which dominate the decay for most $\omega$ in potentials with parity symmetry.
Thus as oscillons age and $\omega$ increases toward $m$, the oscillon radiates more and more slowly.
However, the oscillon is not infinitely long-lived: a geometrical consequence of living in three dimensions
is that an oscillon must die at a frequency $\omega_\te{crit} < m$ (see e.g., Refs.~\cite{Fodor:2009kf,Fodor:2019ftc,Cyncynates:2021rtf}).
An oscillon therefore typically spends the majority of its lifetime at a frequency close to $\omega_\te{crit}$.

The mechanism of radiation (which is discussed in detail in Refs.~\cite{Fodor:2019ftc,Zhang:2020bec,Cyncynates:2021rtf}) does not play an important role in our analysis, so we may simply characterize the decay rate of the oscillon by its instantaneous lifetime:
\begin{align}
    T_\te{inst}(\omega)&= \f{E_\te{osc}(\omega)}{P_\te{rad}(\omega)}\equiv \f{1}{\Gamma_\te{inst}(\omega)},
\end{align}
where $E_\te{osc}$ is the total bound energy in the oscillon and $P_\te{rad}$ is the power radiated by the oscillon, all measured while the oscillon is at a frequency $\omega$.
[The precise values of $E_\mathrm{osc}(\omega)$ and $P_\mathrm{rad}(\omega)$ may be calculated using the software package \href{https://github.com/SimpleOscillon/Code}{\faGithub\hspace{0.1cm}\textsf{Simple Oscillon}}.]

As shown in earlier sections, friendly axions admit oscillon solutions too, although only the short axion is likely to form them.
One may gain insight into the dynamics of short oscillons by taking a spherically symmetric ansatz
with a fixed radial profile, $\phi_S(t, r) = \Phi_S(t) R_S(r)$, in the background of a homogeneous
long axion, $\phi_L(t, r) = \Phi_L(t)$.\footnote{
    Note that $\phi_S(t, r)$ is in general not separable, since as the oscillon radiates energy it adjusts its radial profile, but neglecting such changes is sufficient on timescales much shorter than the oscillon lifetime.
    Moreover, as demonstrated below, the long axion supplies exactly enough energy to drive the short oscillon at a fixed amplitude via autoresonance, thus maintaining its radial profile.
}
To be self-consistent, we work in the limit where the short axion does not backreact onto the long axion, i.e., $f_S \ll f_L$.
This approximation neglects the effects of radiation, which we include via an artificial damping term with coefficient $\Gamma_\mathrm{inst}$.
In the small $\Phi_L/f_L$ limit, keeping only linear terms,
\begin{subequations}\label{eq:oscillonEffectiveEOM}
\begin{align}
    \ddot{\Phi}_L
    + \f{3}{2t}\dot{\Phi}_L
    &= - m^2\mu^2\Phi_L,
    \\
    \ddot{\Phi}_S
    + \p{\f{3}{2t} + \Gamma_\te{inst}} \dot{\Phi}_S
    &= - V_0'(\Phi_S) - V_1'(\Phi_S)\f{\Phi_L}{f_L}.
\end{align}
\end{subequations}
Here $V_0$ and $V_1$ are effective potentials derived by integrating out $R_S$.\footnote{
    To integrate out the radial profile, one substitutes $\phi_S = \Phi_S(t)R_S(r)$ into the action and integrates over space. The qualitative features of the resulting effective action for $\Phi_S$ are insensitive to the choice of $R_S(r)$ so long as it is monotonically decreasing and changes over a radial length scale of at least $1/m$.
}
Ultimately, the precise form of these potentials is unimportant, their salient feature being that they possess attractive nonlinearities.
These equations are thus precisely in the same form as those for the homogeneous system studied in Ref.~\cite{Cyncynates:2021xzw},
and autoresonance between $\Phi_S$ and $\Phi_L$ therefore occurs for a broad range of initial conditions (although the likelihood that any given patch of space leads to oscillon formation and autoresonance is most easily assessed with simulations).

\begin{figure}[t!]
    \centering
    \includegraphics[width = \columnwidth]{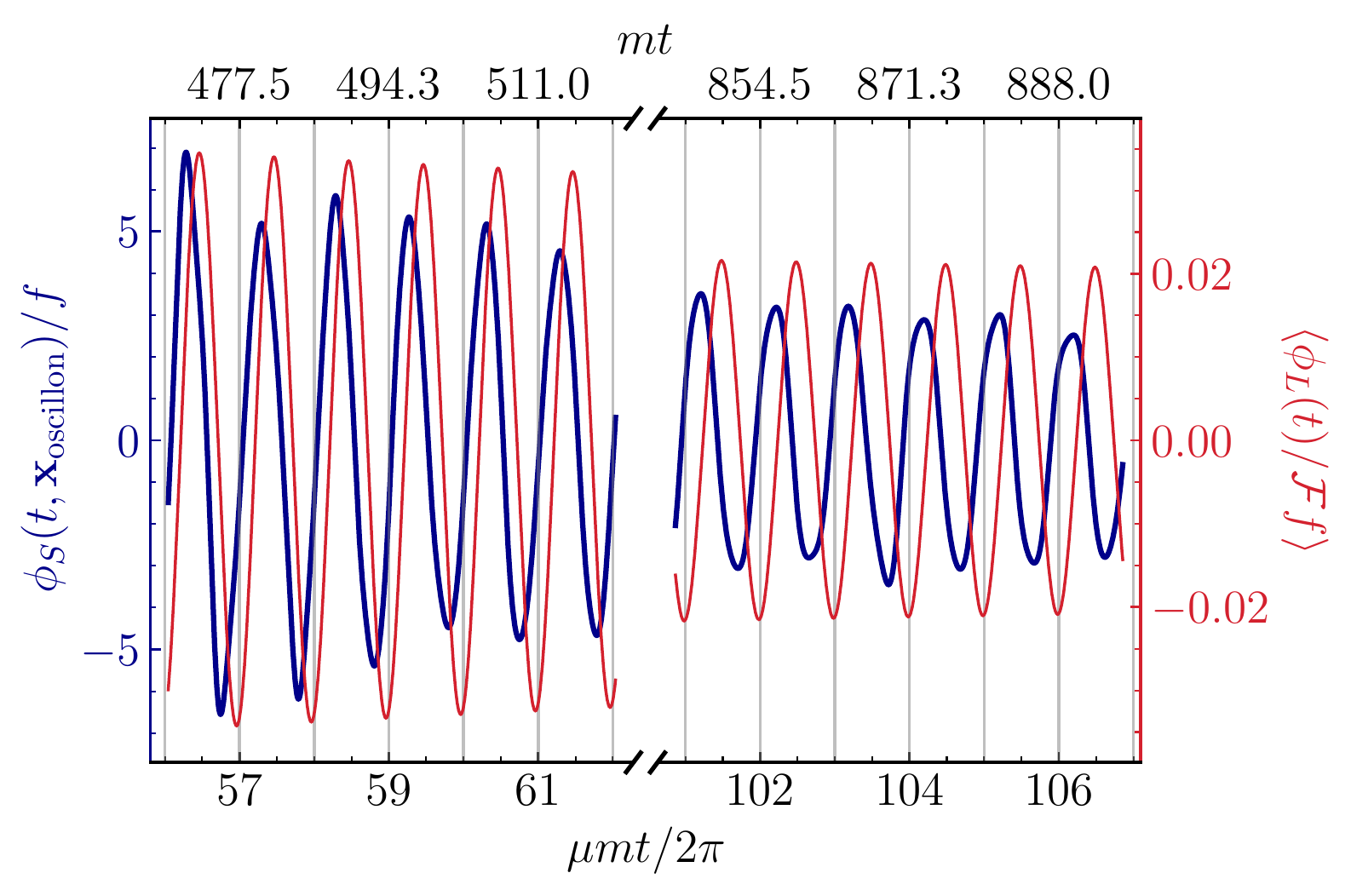}
    \caption{
        Central amplitude of a short-axion oscillon (left vertical axis, blue curves) and
        average amplitude of the long axion (right vertical axis, red curves),
        measured in a cosmological simulation with $N^3 = 512^3$ gridpoints,
        box length $L = \pi / m$, mass ratio $\mu = 0.75$, and decay constant ratio $\calF = 50$.
        Note the break in the horizontal axis between $520 \lesssim m t \lesssim 850$.
    }
    \label{fig:phaseLock}
\end{figure}

\Cref{fig:phaseLock} demonstrates that autoresonance between the short oscillons and the long axion occurs in our full $3+1D$ simulations.
The short-axion oscillon oscillates at the same frequency $\omega = \mu m$ as the driver
$\left\langle \phi_L \right\rangle$, conclusively demonstrating local nonlinear resonance
inside the oscillon.
The phase offset between the short and long axion evolves from roughly $\pi/3$ at early times
(left side of \cref{fig:phaseLock}) to about $\pi/2$ just before oscillon death (right side),
indicating an increasingly efficient energy transfer from the long axion to the short oscillon.
The energy transfer rate peaks when the phase shift reaches $\pi / 2$, and subsequently decreases
as the long axion's amplitude redshifts.
In analogy to homogeneous autoresonance, the long axion can then no longer support the short
axion against its own radiation.

\begin{figure}[t!]
    \centering
    \includegraphics[width = \columnwidth]{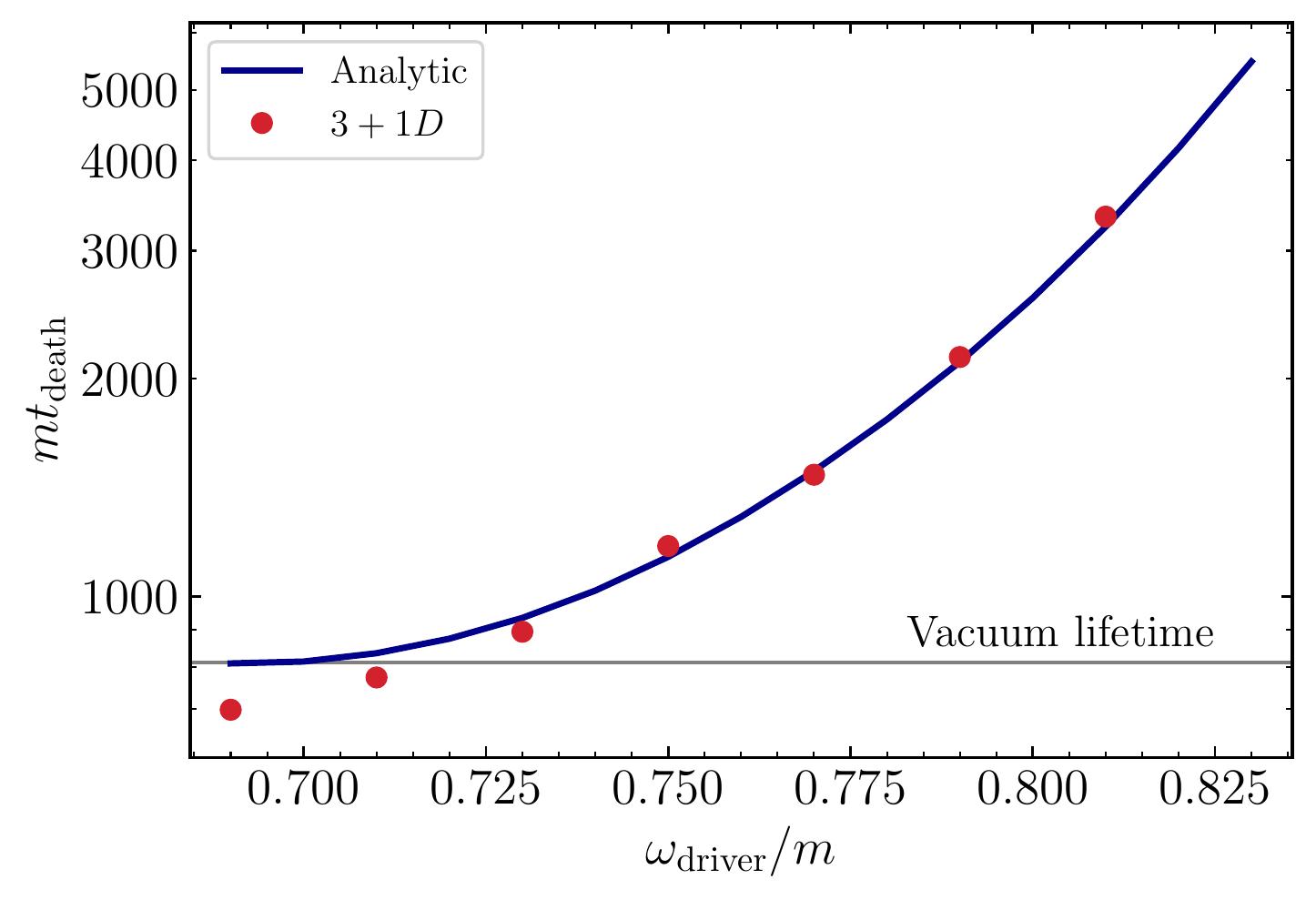}
    \caption{
        Driven oscillon lifetime measured in $3+1D$ cosmological simulations (red dots) versus the
        analytic prediction of \cref{eq:driveLimit} (blue line) at a decay constant ratio
        $\calF = 1000$ (large enough that backreaction is never important).
        The analytic curve only contains one free parameter corresponding to an order 1
        multiplicative constant in \cref{eq:driveLimit} (see \cref{app:oscillons}).
        Note that the driver frequency $\omega_\mathrm{driver}$ simply corresponds to the
        mass of the long axion, $\mu m$.
    }
    \label{fig:lifetime}
\end{figure}

Using the equations of motion \cref{eq:oscillonEffectiveEOM} for the spherically symmetric system, we can solve for the dynamics of the short oscillon and the long driver to arrive at the driven oscillon lifetime $t_\te{death}$ in terms of its instantaneous vacuum lifetime $T_\te{inst}(\mu)$.
The details of this calculation are described in \cref{app:oscillons}, and we summarize the results here.
First, the long driver amplitude must be large enough that it supplies sufficient energy to the oscillon, leading to
\begin{align}
\label{eq:driveLimit}
    m \mu t_\te{death}\approx \left[ m \mu T_\te{inst}(\mu) \right]^{4/3}.
\end{align}
We compare this analytic scaling to simulations in \cref{fig:lifetime}, verifying the $[\mu T_\mathrm{inst}(\mu)]^{4/3}$
dependence in the range $0.73 \lesssim \mu \leq 0.81$.
but spherically symmetric simulations verify the scaling of \cref{eq:driveLimit} out to $\mu = 0.89$.
At lower frequencies the driven oscillon lifetime is shorter than the vacuum lifetime: once
the oscillon falls off autoresonance, it rapidly dumps its energy into the long axion field,
cutting its life short.
Larger values of $\mu$ require longer ($3+1D$) simulation runtimes than our computational resources permit.

\begin{figure}[t!]
    \centering
    \includegraphics[width = \columnwidth]{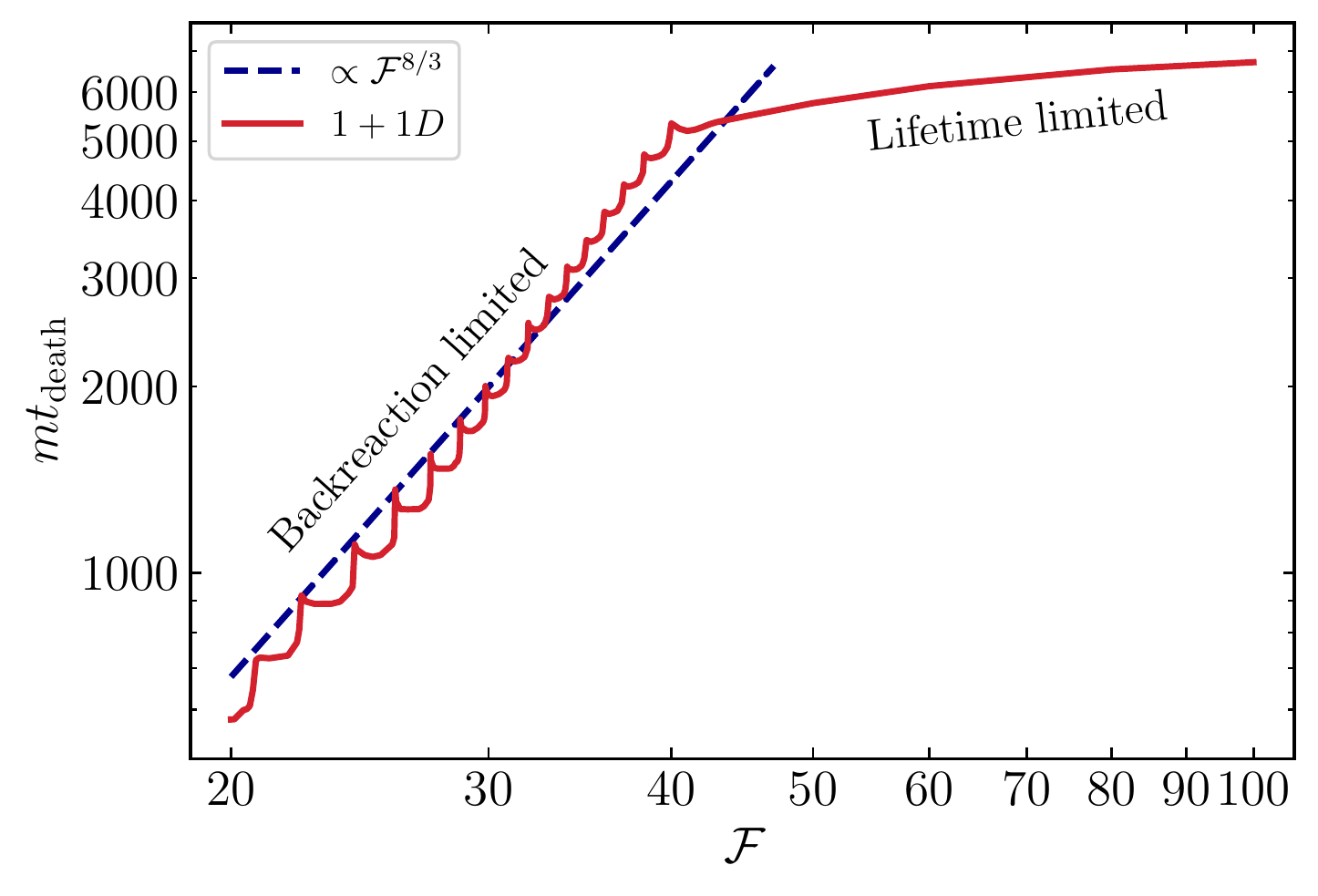}
    \caption{
        Driven oscillon lifetime versus decay constant ratio $\calF$ in spherically symmetric
        simulations with $\mu = 0.85$ (red) with superposed $\calF^{8/3}$ scaling predicted by
        \cref{eq:BRLimit} (blue).
        The wiggles in the red curve are due to the precise phase relationship between the oscillon
        and the driver, a feature that would be washed out in any realistic simulation with randomly
        sampled initial phases.
    }
    \label{fig:Backreaction}
\end{figure}

The oscillon also backreacts on the driver, inducing spatial perturbations.
These fluctuations remain small only until
\begin{align}
\label{eq:BRLimit}
    m \mu t_\te{death}
    \lesssim \calF^{8/3}.
\end{align}
This scaling is demonstrated in \cref{fig:Backreaction}: the spherically symmetric solutions to the full coupled nonlinear wave equations (see Eq.~\ref{eq:OscillonEOM}) exhibit $\calF^{8/3}$ behavior for $\calF \lesssim 40$,
at which point the lifetime saturates the driving limit, \cref{eq:driveLimit}.
Performing a quantitative analysis of the parametric dependence on $\mathcal{F}$ in $3+1D$ simulations would require simulations with larger mass ratio $\mu$, which, as discussed previously, exceed the limits of our computing resources.
However, we do observe qualitative behavior---lifetimes that increase with $\mathcal{F}$ and saturate at the driving limit in \cref{fig:lifetime}---similar to that of \cref{fig:Backreaction} in the set of simulations presented above.

Finally, the oscillon siphons energy from the long axion, depleting the latter's local energy density.
Nearby regions of space then resupply this region with long axion; requiring that this resupply rate exceeds the
depletion rate due to the oscillon leads to the final constraint
\begin{align}
\label{eq:DepletionLimit}
    t_\te{death}\lesssim \calF^2 T_\te{inst}(\mu).
\end{align}
One can check that there is no region of $\calF-T_\mathrm{inst}$ parameter space where \cref{eq:DepletionLimit} dominates the lifetime.
Taken together, the lifetime is
\begin{align}
\label{eq:lifetime}
    m \mu t_\te{death}
    &\approx \min\left( \left[ m \mu T_\te{inst}(\mu) \right]^{4/3}, \calF^{8/3} \right).
\end{align}
While the lifetimes of these driven oscillons are parametrically enhanced relative to their in-vacuum expected lifespans, they are still far too short-lived to be of any cosmological relevance.
Nonetheless, these novel dynamics---interesting in their own right---potentially broaden the class of scalar field theories that admit oscillons surviving into the present day.
We discuss this possibility and associated challenges in App.~\ref{app:generalPotentialLongevity}.

\section{Discussion} \label{sec:discussion}

Nonlinear effects in the early Universe can have a drastic impact on the late-time distribution of energy in dark sectors.
The ``friendly'' axion system of Ref.~\cite{Cyncynates:2021xzw} provides a concrete example model, where nonlinearities dominate the dynamics of both the background and fluctuations and have important consequences for direct detection experiments.
In this paper, we numerically evolved the full system, showing that large axion fluctuations---in particular the nucleation of oscillons---work to equilibrate the relic densities of the two axions for a moderate ratio of the heavy axion's decay constant to that of the light one
($6 \lesssim \calF \lesssim 20$).
For smaller decay constant ratios $\calF \lesssim 6$, spatial fluctuations have a negligible effect
on the dynamics, and we recover the results of homogeneous computations from
Ref.~\cite{Cyncynates:2021xzw}.  At larger ratios $\calF \gg 20$, oscillon nucleation prevents the heavy axion from ever
attaining a substantial abundance.

The novel dynamics in the intermediate-$\calF$ regime position friendly axions to be positively identified
as two-component dark matter by direct detection experiments.
The lighter axion's abundance is reduced by no more than a factor of about two, in sharp contrast to expectations based on homogeneous approximations in which its abundance would be parametrically depleted~\cite{Cyncynates:2021xzw}.
The heavier axion's abundance (and therefore detection prospects) is still parametrically enhanced (by a factor of $\approx \calF^2 / 2$), but only at a moderate cost to the visibility of the lighter axion.
Many upcoming axion direct detection experiments~\cite{Brouwer:2022bwo,Alesini:2017ifp,Stern:2016bbw,DMRadio:2022pkf} would potentially be sensitive to \textit{both} axions in a friendly pair having masses within the experiment's sensitivity band.
Direct detection of axion dark matter with a decay constant substantially smaller than that
expected in standard misalignment scenarios should prompt a search for a second, more weakly
coupled axion at a nearby mass.

We also computed the stochastic gravitational wave background produced by oscillon nucleation in a friendly axion sector.
If friendly axions compose all of the dark matter, the present-day strain is well out of reach of near-future gravitational wave experiments, but the cosmic microwave background polarization does constrain (and in the future may probe) hyperlight friendly pairs making up a subcomponent of dark matter.
Density and vector perturbations are also produced in these scenarios; their effect on the CMB
(and other cosmological observables) is less straightforward to evaluate, but they may well provide
even more stringent constraints than just the (as-yet unobserved) primordial $B$-mode polarization.

Finally, for $\calF \gtrsim 20$, although autoresonance is quenched by oscillon production (preventing the axions' energies from equalizing) our simulations demonstrate that short-axion oscillons produced in the early Universe are driven by the long axion background, parametrically extending their lifetimes.
For the specific friendly axion potential studied here (\cref{eq:twoAxionPotential}), driven oscillons can live about an order of magnitude longer than their undriven counterparts.
Though they are still not long-lived enough to be astrophysically relevant, even at the lightest possible axion masses, this may not be the case for other scalar potentials.
Similar dynamics in other coupled axion theories may lead to driven oscillons that could naturally live until the present day, with numerous possible observational signatures including gravitational lensing~\cite{VanTilburg:2018ykj}, optical lensing~\cite{Prabhu:2020pzm}, and electromagnetic bursts~\cite{Prabhu:2020yif,Buckley:2020fmh,Amin:2020vja,Amin:2021tnq}.

Our results are qualitatively insensitive to the amplitude of the initial primordial curvature perturbations.
However, the size does determine the
precise minimum and maximum decay constant ratios for which the two axion energy densities equalize.
For simplicity, we used a scale-invariant initial power spectrum with magnitude set by the \textit{Planck} measurements at CMB scales, but in reality adiabatic fluctuations are red-tilted on large scales and are much less constrained on smaller scales.
If there is less initial power at small scales, the time to nonlinearity and oscillon formation increases, allowing friendly pairs with larger decay constant ratios $\calF \gg 20$ to achieve equipartition.
We also note that whether the initial axion perturbations are adiabatic as above or seeded directly in the axion field (i.e., isocurvature perturbations) does not measurably affect any of our results, as corroborated by simulations with purely isocurvature initial conditions.

The string axiverse is a rare example of a low-energy signature of quantum gravity,
most of whose novel predictions reside at the grand-unified or string scales, far outside experimental reach.
In general, an axiverse can comprise a multitude of light, coupled axions; this work provides further
evidence of the outsized impact nonlinear effects and interactions have on the phenomenology of the axiverse.
The friendly model considered here, for which nonperturbative dynamics revise predictions by multiple orders of magnitude, is only a prototypical example; further work is necessary to understand the phenomenology of fully realistic axiverses and the critical role played by nonlinear dynamics.

\begin{acknowledgments}
We thank Masha Baryakhtar, Davide Racco, and the anonymous referee for thoughtful feedback on this manuscript, Savas Dimopoulos for useful discussions on ``oscillon longevity centers,'' and Dmitriy Zhigunov for helpful conversations about strongly nonlinear fields.
D.C.\ is grateful for the support of the Stanford Institute for Theoretical Physics (SITP), the National Science Foundation under Grant No.\ PHY-2014215, and the Gordon and Betty Moore Foundation under Grant No.\ GBMF7946. O.S.\ is supported by a DARE fellowship from the Vice Provost for Graduate Education at Stanford University. J.O.T.\ is supported by the ARCS Foundation, and is thankful to the Perimeter Institute for its hospitality during the final stages of completing this manuscript.
Z.J.W.\ is supported by the Department of Physics and the College of Arts and Sciences at the
University of Washington.
This work used the Extreme Science and Engineering Discovery Environment (XSEDE)~\cite{xsede}, which
is supported by National Science Foundation Grant No.~ACI-1548562; simulations were run through allocation No.~TG-PHY200037 on the Anvil cluster at Purdue University, Bridges-2 at the Pittsburgh Supercomputing Center which is supported by NSF Grant No.~ACI-1928147, and
Expanse at the San Diego Supercomputer Center. This research was supported in part by Perimeter Institute for Theoretical Physics. Research at Perimeter Institute is supported by the Government of Canada through the Department of Innovation, Science and Economic Development and by the Province of Ontario through the Ministry of Research, Innovation and Science.

Simulations in this work were implemented with \textsf{pystella}~\cite{pystella}, which is available at
\href{https://github.com/zachjweiner/pystella}{github.com/zachjweiner/pystella} and makes use of
the Python packages \textsf{PyOpenCL}~\cite{kloeckner_pycuda_2012},
\textsf{Loopy}~\cite{kloeckner_loopy_2014}, \textsf{mpi4py}~\cite{DALCIN2008655,DALCIN20051108},
\textsf{mpi4py-fft}~\cite{jpdc_fft}, and \textsf{NumPy}~\cite{Harris:2020xlr}.
This work also made use of the packages \textsf{SciPy}~\cite{Virtanen:2019joe}, \textsf{matplotlib}~\cite{Hunter:2007ouj}, \textsf{SymPy}~\cite{Meurer:2017yhf}, and
\textsf{CMasher}~\cite{cmasher}.

\end{acknowledgments}

\appendix

\section{Equations of motion and numerical implementation}\label{app:numerical-details}

For completeness, we enumerate the evolution equations as implemented in simulations.
We use a conformal-time, ``mostly plus'' FLRW metric in the conformal Newtonian gauge with line
element
\begin{align}
\begin{split}
    \mathrm{d} s^2
    &= - a(\tau)^2 \left[ 1 + 2 \Phi(\tau, \mathbf{x}) \right] \mathrm{d} \tau^2
    \\ &\hphantom{ {}={} }
        + a(\tau)^2 \left[
            \left\{ 1 - 2 \Phi(\tau, \mathbf{x}) \right\} \delta_{ij}
            + h_{ij}(\tau, \mathbf{x})
        \right]
        \mathrm{d} x^i \mathrm{d} x^j.
\end{split}
\end{align}
Here $\Phi$ is the Newtonian potential and $h_{ij}$ the transverse ($\partial_i h_{ij} = 0$) and
traceless ($\delta^{ij} h_{ij} = 0$) tensor perturbation.
We neglect scalar anisotropic stress and vector perturbations.
We define the comoving Hubble parameter $\mathcal{H} \equiv \partial_\tau a / a$ (while the standard
Hubble parameter is $H \equiv \partial_t a / a = \mathcal{H} / a$).

Well before matter-radiation equality, the axions make a negligible contribution to the expansion of
the Universe; the solution to the Friedmann equations in a radiation Universe is
$a(\tau) / a(\tau_m) = \tau / \tau_m$.
We take $\tau_m = m^{-1}$ so that the scale factor is normalized to the time when $H = m$.
Because tensor perturbations (i.e., gravitational waves) from inflation are very small and their
subsequent production by axion production is suppressed by $(f_L / \Mpl)^2$, we neglect
their backreaction onto the axion fields.
For the same reason, we neglect the contribution of the axions to the Newtonian potential.

\subsection{Equations of motion}

The Euler-Lagrange equation for the axions reads
\begin{align}
\begin{split}\label{eqn:phi-I-eom}
    \partial_\tau^2 \phi_I
    &= - \left[
            2 \mathcal{H}(\tau)
            - 4 \partial_\tau \Phi
        \right]
        \partial_\tau \phi_I
        + \left( 1 + 4 \Phi \right) \partial_i \partial_i \phi_I
    \\ &\hphantom{ {}={} }
        - a(\tau)^2 \left( 1 + 2 \Phi \right) \frac{\partial V}{\partial \phi_I},
\end{split}
\end{align}
where the potential $V$ is defined in \cref{eq:twoAxionPotentialPhi} and $I = S$ or $L$.
(We use repeated Latin indices $i$, $j$, $k$, etc., to denote spatial components that are contracted
with the Kronecker delta regardless of their placement.)
In a Universe dominated by a single fluid with equation of state $w$ and a sound speed
$c_s^2 \equiv \delta P / \delta \rho$, the Einstein equations for $\Phi$ may be rearranged into
\begin{align}
\begin{split}\label{eqn:newtonian-eom}
    \partial_\tau^2 \Phi
    &= - (2 + 3 c_s^2) \mathcal{H} \partial_\tau \Phi
        - \mathcal{H} \partial_\tau \Phi
    \\ &\hphantom{ {}={} }
        - 3 \left( c_s^2 - w \right) \mathcal{H}^2 \Phi
        + c_s^2 \partial_i \partial_i \Phi,
\end{split}
\end{align}
where both $w$ and $c_s^2$ are $1/3$ in the radiation-dominated era.
Finally, the tensor perturbations evolve according to
\begin{align}\label{eqn:gw-eom}
    \partial_\tau^2 h_{ij} + 2 \mathcal{H} \partial_\tau h_{ij} - \partial_k \partial_k h_{ij}
    &= \frac{2 a(\tau)^2}{\Mpl^2} \Pi^{i}_{\hphantom{i}j}.
\end{align}

Gravitational waves are sourced by the transverse and traceless anisotropic stress tensor
$\Pi_{ij}$ whose Fourier modes are given in terms of the full stress tensor
$T_{ij}$ by
\begin{align}
    \Pi_{ij}(\tau, \mathbf{k})
    &= \left(
            P_{il}(\mathbf{k}) P_{jm}(\mathbf{k})
            - \frac{1}{2} P_{ij}(\mathbf{k}) P_{lm}(\mathbf{k})
        \right) T_{lm}(\tau, \mathbf{k})
\end{align}
with
\begin{align}
    P_{ij}(\mathbf{k})
    &= \delta_{ij} - \frac{k_i k_j}{k^2}.
\end{align}
Unlike the Newtonian potential, which is dominated by the standard adiabatic perturbations in the radiation fluid, the gravitational waves are only sourced by the axions.
However, the sourced gravitational waves are also suppressed by $(f_L / \Mpl)^2$, for
which reason it is entirely sufficient to evaluate the stress tensor as if in a homogeneous FLRW Universe:
\begin{align}\label{eqn:full-stress-tensor}
\begin{split}
    T^i_{\hphantom{i}j}
    &= \sum_I \partial_i \phi_I \partial_j \phi_I
    \\ &\hphantom{{}={}}
        - \delta^i_{\hphantom{i}j}
        \left(
            \frac{1}{2} \sum_I \partial_\mu \phi_I \partial^\mu \phi_I
            + V(\phi_S,\phi_L)
        \right).
\end{split}
\end{align}
The final term purely contributes to the trace of the stress-energy tensor and may be dropped when computing the metric tensor perturbations.
Because the backreaction of gravitational waves on the axions is negligible, we may simply ignore any
initial amplitude generated during inflation and consider the evolution of \cref{eqn:gw-eom} from
zero initial conditions.

\subsection{Initial conditions}

Imposing that $\Phi$ was frozen outside of the horizon ($k \tau \ll 1$) to its primordial value
generated during inflation sets $\Phi(\tau \ll 1/k, \mathbf{k}) = \Phi_0(\mathbf{k})$
and $\partial_\tau \Phi(\tau \ll 1/k, \mathbf{k}) = 0$.
In this case, we find the solution
\begin{align}\label{eqn:Phi-analytic-solution}
    \Phi(\tau, \mathbf{k})
    &= \Phi_0(\mathbf{k}) \frac{\sin y - y \cos y}{y^3},
\end{align}
where $y \equiv \sqrt{w} k \tau$.
The primordial curvature perturbation is characterized by a dimensionless power spectrum
$\Delta_\Phi^2(k)$ defined by
\begin{align}\label{eqn:def-dimless-power-spectrum-Phi}
    \frac{k^3}{2 \pi^2}
    \left\langle \Phi_0(\mathbf{k}_1) \Phi_0(\mathbf{k}_2) \right\rangle
    &= (2 \pi)^3 \delta(\mathbf{k}_1 + \mathbf{k}_2)
        \Delta_\Phi^2(k),
\end{align}
which, for standard slow-roll inflation, is nearly scale invariant with amplitude $\sim 10^{-9}$~\cite{Planck:2018vyg}.
To avoid specifying the mass scale of the problem, we neglect the small spectral tilt
and simply take a scale-invariant spectrum.

We begin the simulations when $H = m$ (at $\tau = \tau_1$), solving the linearized system of equations from an early time when all relevant wave numbers are well outside the horizon (i.e., when $k \ll a H$) to determine initial conditions.
The axions are initialized as Gaussian-random fields with mean field value and velocity at $\tau_1$ set according to the solution to the homogeneous system.
The fluctuations are set in Fourier space to match the linearized solution $\phi_{I, k}(\tau_1)$, multiplied by a Gaussian-random complex number (normalized such that the mean squared modulus is unity).
For each axion and wave vector $\mathbf{k}$ on the grid, we therefore set
\begin{align}
    \phi_I(\tau_1, \mathbf{k})
    &= \phi_{I, k}(\tau_1) \sqrt{- \ln U_1(\mathbf{k})} e^{2 \pi i U_2(\mathbf{k})} \\
    \partial_\tau \phi_I(\tau_1, \mathbf{k})
    &= \partial_\tau \phi_{I, k}(\tau_1) \sqrt{- \ln U_1(\mathbf{k})} e^{2 \pi i U_2(\mathbf{k})},
\end{align}
where $U_1(\mathbf{k})$ and $U_2(\mathbf{k})$ are random variates between 0 and 1, each of which are the same for both axions.
The Newtonian potential is initialized analogously (and with random numbers for each wave vector matching those for the axions) using the analytic solution \cref{eqn:Phi-analytic-solution}.

We separately implemented simulations that neglect metric perturbations, instead initializing the
axion with a scale-invariant spectrum of isocurvature fluctuations (with amplitude comparable to
that of the adiabatic initial conditions).
At the level of the analysis performed in \cref{sec:results}, the results are effectively unchanged,
since the initial size and early evolution of fluctuations are important only insofar as they affect
the time to nonlinearity.
Nonlinear interactions quench autoresonance and distribute power across length scales, largely
washing out any detailed features of the initial conditions.
Since amplification (due to parametric resonance) does not begin immediately after modes start
oscillating, quantitative results retain only a small, logarithmic dependence on initial conditions.

\subsection{Gravitational wave backgrounds}\label{app:gravitational-waves}

Gravitational waves carry an effective energy density which, deep inside the horizon (i.e.,
$k \gg \mathcal{H}$), is~\cite{Abramo:1997hu,Brandenberger:2018fte,Clarke:2020bil}
\begin{align}
    \rho_\mathrm{GW}(\tau)
    = \frac{\Mpl^2}{4 a^2}
        \left\langle
            \overbar{\partial_\tau h_{ij} \partial_\tau h_{ij}}
        \right\rangle,
    \label{eqn:rho-gw-subhorizon}
\end{align}
where the bar denotes a time average (i.e., over oscillations).
The relic abundance of gravitational waves today per logarithmic wave number is
\begin{align}\label{eqn:omega-gw-spectrum-def}
    \Omega_\mathrm{GW}(\tau, k)
	= \frac{1}{\bar{\rho}(\tau)} \dd{\rho_\mathrm{GW}(\tau)}{\ln k}.
\end{align}
Substituting the inverse Fourier transform of $h_{ij}$ into \cref{eqn:rho-gw-subhorizon} permits
rewriting \cref{eqn:omega-gw-spectrum-def} in terms of the dimensionless power spectrum of
$\partial_\tau h_{i j}$ (defined in analogy to \cref{eqn:def-dimless-power-spectrum-Phi}) as
\begin{align}\label{eqn:omega-gw-spectrum-ito-power-spectrum}
    \Omega_\mathrm{GW}(\tau, k)
	= \frac{1}{12 \mathcal{H}(\tau)^2}
        \sum_{i, j} \Delta^2_{\partial_\tau h_{ij}}(\tau, k),
\end{align}
after plugging in $\bar{\rho}(\tau) = 3 \Mpl^2 H(\tau)^2$.
The spectrum evaluated in the early Universe is related to that at the present day (at $\tau_0$) by
the transfer function
\begin{align}
    \frac{
        \Omega_{\mathrm{GW}}(\tau_0, k) h^2
    }{
        \Omega_{\mathrm{GW}}(\tau, k)
    }
    &= \Omega_{\mathrm{rad}}(\tau_0) h^2
        \frac{g_{\star}(\tau)}{g_{\star}(\tau_0)}
        \left( \frac{g_{\star S}(\tau)}{g_{\star S}(\tau_0)} \right)^{-4/3},
    \label{eqn:gw-amplitude-transfer-function}
\end{align}
and would be observed at present-day frequencies related to the wave number $k$ by
\begin{align}
\begin{split}
    f_\mathrm{GW}
    &= \frac{k / 2 \pi a(\tau)}{\sqrt{H(\tau) \Mpl}}
        \left[
            \Omega_{\mathrm{rad}}(\tau_0)
            H_0^2 \Mpl^2
        \right]^{1/4}
    \\ &\hphantom{{}={} }
        \times
        \left( \frac{g_{\star}(\tau)}{g_{\star}(\tau_0)} \right)^{1/4}
        \left( \frac{g_{\star S}(\tau)}{g_{\star S}(\tau_0)} \right)^{-1/3}.
\end{split}
\end{align}
Here $g_{\star}$ and $g_{\star S}$ are the numbers of relativistic degrees of freedom in energy and
entropy density, respectively.
Note that the present-day abundance of radiation is
$\Omega_{\mathrm{rad}}(\tau_0) h^2 \approx 4.2 \times 10^{-5}$~\cite{Planck:2018vyg} and that
$H_0 / h \equiv 100 \, \mathrm{km} \, \mathrm{s}^{-1} / \mathrm{Mpc} \approx 3.24 \times 10^{-18} \, \mathrm{Hz}$.

\subsection{Numerical implementation}

We discretize the evolution equations, \cref{eqn:phi-I-eom,eqn:newtonian-eom,eqn:gw-eom},
onto a three dimensional, regularly spaced grid with periodic boundary conditions.
Following Refs.~\cite{Adshead:2019igv,Adshead:2019lbr,Weiner:2020sxn}, we evolve the gravitational
wave equation of motion (\cref{eqn:gw-eom}) under the replacement of the transverse-traceless
stress tensor (which is only easily calculated in Fourier space) with the full stress tensor
(\cref{eqn:full-stress-tensor}).
The transverse-traceless projection is instead performed on the tensor field itself only when
outputting gravitational wave spectra, drastically reducing the required number of fast Fourier
Transforms (which in distributed-memory contexts are a major bottleneck).
For similar reasons, rather than use the analytic solution \cref{eqn:Phi-analytic-solution} for the
Newtonian potential (which requires forward and inverse Fourier transforms at each step), we simply
evolve \cref{eqn:newtonian-eom} in position space.

Spatial derivatives are approximated with fourth-order centered differencing, while integration in time is implemented with a ``low-storage,'' fourth-order Runge-Kutta method~\cite{carpenter1994fourth}.
We have verified that the results are insensitive to the precise choice of box length $L$ and are consistent with simulations with the same physical volume but $1280^3$ and $1536^3$ gridpoints (compared to the $1024^3$ used for main results), as well as at the same resolution with a box length $1.25$ and $1.5$ times larger.
We have also checked for a representative case that results at a fixed resolution and volume are qualitatively
insensitive to the the random seed used in generating stochastic initial conditions.

\section{Details of driven oscillons} \label{app:oscillons}

In \cref{sec:drivenOscillons}, we demonstrated that short-axion oscillons remain in autoresonance with the nearly homogeneous long axion, often extending their lifetime well beyond the in-vacuum expectation.
The physics of this localized autoresonance is mostly captured by the effective equations of motion (\cref{eq:oscillonEffectiveEOM})
obtained by integrating out the spatial profile of the oscillon, demonstrating that the short oscillon is well approximated as a one dimensional driven nonlinear oscillator.
However, \cref{eq:oscillonEffectiveEOM} fails to capture some important effects that ultimately limit the lifetime of the driven oscillon, namely \cref{eq:driveLimit,eq:BRLimit,eq:DepletionLimit}.
In what follows, we derive the lifetime bounds in \cref{eq:driveLimit,eq:BRLimit,eq:DepletionLimit} in \cref{app:lifetimeBounds} and discuss the possibility of cosmologically long-lived oscillons in more general multiaxion potentials in \cref{app:generalPotentialLongevity}.

\subsection{Lifetime bounds}
\label{app:lifetimeBounds}

Most long-lived oscillons are well approximated by a single-harmonic ansatz, so $\Phi_S$ oscillates predominantly at a single frequency.
On autoresonance, this frequency approximately matches the long-axion mass ($\mu m$) up to transient oscillations around this stable point.
Thus on autoresonance we may approximate
\begin{align}
    \Phi_S &\approx f_S\Theta_S^0\sin(m\mu t + \delta),\\
    \label{eq:oscillonLongAmplitude}
    \Phi_L &\approx f_L\Theta_L^0\p{\f{t}{t_0}}^{-3/4}\sin(m\mu t),
\end{align}
where $\delta$ is a phase-offset.
While $\delta$ does have a time dependence, as discussed in App.~B.3 of~\cite{Cyncynates:2021xzw}, it is slow compared to the oscillatory timescale and thus we can approximate $\delta$ as a constant.
The power transferred from the long to the short axion is
\begin{align}
    \f{P_{L\to S}}{V_\te{osc}}&=\p{\dot \Phi_S\partial_{\Phi_S} - \dot \Phi_L\partial_{\Phi_L}} V_\te{int},
\end{align}
where for our purposes we may approximate the interaction potential by a mass-mixing term,
\begin{align}
    V_\te{int} \sim m^2 f_S \Phi_S\f{\Phi_L}{f_L}.
\end{align}
Thus, the time-averaged power transfer is
\begin{align}
    \gen{P_{L\to S}} \approx -m\mu E_\te{osc} \f{\Theta_L^0}{\Theta_S^0}\p{\f{t}{t_0}}^{-3/4}\sin\delta,
\end{align}
where we've taken $m^2 f_S^2 (\Theta_S^0)^2 V_\te{osc}\approx E_\te{osc}$.
As shown in Ref.~\cite{Cyncynates:2021xzw}, the phase $\delta$ tends toward $-\pi/2$ toward the end of autoresonance, and thus we take $\delta = -\pi/2$ when calculating the driven oscillon lifetime.
The oscillon is supported by the driver when its radiated power $P_\mathrm{rad}$ is smaller than the maximum power transfer from the long axion $\gen{P_{L\to S}}$, from which we obtain the approximate time of death $t_\mathrm{death}$,
\begin{align}
    \label{eq:drivenLifetimeApp}
    m\mu t_\te{death} \approx  \p{m\mu\f{\Theta_L^0}{\Theta_S^0}T_\te{inst}(\mu)}^{4/3},
\end{align}
taking $t_0 \approx 1/m\mu$.
Here $T_\mathrm{inst}(\omega) = E_\te{osc}(\omega)/P_\te{rad}(\omega)$ is the instantaneous oscillon lifetime at the frequency $\omega$, which may be longer or shorter than the in-vacuum oscillon lifetime.
Provided that the driving frequency $m\mu$ corresponds to a slow-decaying part of the oscillon lifecycle, this result shows that the driven oscillon lifetime is parametrically enhanced relative to its vacuum lifetime.

So far, we have neglected the backreaction of $\theta_S$ onto $\theta_L$ in order to approximate $\theta_L$ as a homogeneous field.
While the details of backreaction are complicated, it is simple to obtain an estimate for when $\theta_S$ causes ${\cal O}(1)$ fluctuations in $\theta_L$ which then terminate the autoresonance.
To make this estimate, we observe in the equations of motion,
\begin{subequations}
\label{eq:OscillonEOM}
\begin{align}
\label{eq:longOscillonEOM}
    \square\theta_L + m^2 \calF^{-2}\sin(\theta_S + \theta_L) + m^2 \mu^2\sin\theta_L
    &= 0 \\
    \square\theta_S + m^2 \mu^2\sin(\theta_S + \theta_L)
    &= 0,
\end{align}
\end{subequations}
that $\theta_L$'s evolution depends on $\theta_S$ multiplied by $\calF^{-2}$.
Comparing this term to the final term in \cref{eq:longOscillonEOM}, we see that backreaction becomes important when
\begin{align}
    \label{eq:backreactionLifetime}
    \theta_L \sim \calF^{-2}\theta_S.
\end{align}
Assuming that $\theta_S$ has a constant $\mathcal{O}(1)$ amplitude, as inside an oscillon, and that $\theta_L$ decays like cold matter as in \cref{eq:oscillonLongAmplitude}, we find that spatial fluctuations in $\theta_L$ become order one at the time
\begin{align}
    m\mu t_\te{death} \sim \p{\f{\Theta_L^0}{\Theta_S^0}}^{4/3}\calF^{8/3}.
\end{align}
After this time, the long axion no longer serves as a good driver and quickly dephases with the short axion oscillon, siphoning its energy and causing its rapid death.
We plot this predicted $\calF^{8/3}$ scaling versus the observed lifetime in a spherically symmetric driven oscillon simulation in \cref{fig:Backreaction}.
The $\calF^{8/3}$ scaling at small $\calF$ is eventually replaced by approximately constant scaling at larger $\calF$ when the lifetime bound \cref{eq:drivenLifetimeApp} takes over.

As the oscillon siphons energy from the long axion, it depletes the energy density in its local environment of volume
\begin{align}
    V_\te{dep} \sim \f{P_\te{rad} t}{\rho_L}.
\end{align}
The surrounding long axion flows into the depleted volume at a velocity which we approximate assuming the inner depleted region is diminished by ${\cal O}(1)$:
\begin{align}
    v
    = \frac{\mathrm{d} \omega}{\mathrm{d} k}
    = \f{k}{m_L} \sim \f{2\pi}{m \mu R_\te{dep}}.
\end{align}
where $V_\te{dep} = 4\pi/3 R_\te{dep}^3$.
Using this velocity, we calculate the power flowing from the environment into the depleted region:
\begin{align}
    P_{\te{env}\to\te{dep}} = 4\pi R_\te{dep}^2 v\rho_L.
\end{align}
Solving for the time at which $P_{\te{env}\to\te{dep}} = P_\te{rad}$, we find the following upper bound on the oscillon lifetime
\begin{align}
    m\mu t_\te{death}\lesssim\p{\f{\Theta_L^0}{\Theta_S^0}}^2\calF^2 m T_\te{inst}(\mu).
\end{align}

\subsection{General potentials and cosmological longevity}
\label{app:generalPotentialLongevity}

For the simplest two-axion potentials, the lifetime of driven oscillons is still far shorter than the age of the Universe.
Generalizations of \cref{eq:twoAxionPotential} of the form
\begin{align}
    V(\ths,\thl) = V_S(\ths + \thl) + V_L(\thl),
\end{align}
could plausibly be constructed so that driven oscillons survive into the present day.
The ``vacuum lifetime'' of the short oscillon $T_\mathrm{inst}(\mu)$ is then the instantaneous lifetime of single-field oscillons in the $V_S(\ths)$ potential.

During hierarchical structure formation, a driven short-axion oscillon finds itself in long-axion halos of increasing size, which could make it energetically possible for the oscillon to live indefinitely, although there are some significant uncertainties.
A stationary oscillon in a static long axion halo would eventually deplete its local environment of long axion, starving itself of the energy it needs to survive.
In a realistic galactic halo, however, the region inside the oscillon is constantly being replenished via the virial motion of the long axion.
To take advantage of this energy source, the oscillon must also adjust its phase to match that of the driver over one virial timescale; it is not clear whether particularly long-lived oscillons can perform this phase alignment without some efficient dissipation mechanism (such as dynamical friction due to photon radiation).
We thus leave the question of cosmologically long-lived driven oscillons to future work.

\bibliography{bibliography}

\end{document}